\documentclass[fleqn,usenatbib]{mnras}

\usepackage{newtxtext,newtxmath}

\usepackage[T1]{fontenc}
\usepackage{ae,aecompl}

\usepackage{graphicx}	
\usepackage{amsmath}	
\usepackage{amssymb}	
\usepackage{pdflscape}	

\newcommand{\patternnumber}{960 }
\newcommand{\starnumber}{611 }
\newcommand{\totalnumber}{2085 }
\newcommand{\shortcadencedatanumber}{97 }
\newcommand{\hybridsnumber}{124 }
\newcommand{\hybridsshortcadencenumber}{10}
\newcommand{\gdor}{$\gamma$\,Dor }
\newcommand{\cpd}{$\mathrm{d^{-1}}$}
\newcommand{\frot}{$f_\mathrm{rot}$ }
\newcommand{\numbersurfacemodulation}{58 }
\newcommand{\numberfastsplitting}{11 }
\newcommand{\numberossbyrmode}{110 }

\title[Period spacings of \starnumber $\gamma$ Dor stars]{Gravity-mode period spacings and near-core rotation rates of \starnumber $\gamma$ Doradus stars with \textit{Kepler}}

\author[G. Li et al.]{
Gang Li$^{1,3\thanks{E-mail: gali8292@uni.sydney.edu.au}}$,
Timothy Van Reeth$^{1,2,3}$,
Timothy R. Bedding$^{1,3\thanks{E-mail: tim.bedding@sydney.edu.au}}$,
Simon J. Murphy$^{1,3}$,
\newauthor{Victoria Antoci$^3$, Rhita-Maria Ouazzani$^4$, Nicholas H. Barbara$^{1,3}$}
\\
$^1$Sydney Institute for Astronomy (SIfA), School of Physics, 2006 University of Sydney, Australia\\
$^2$Institute of Astronomy, KU Leuven, Celestijnenlaan 200D, B-3001 Leuven, Belgium\\
$^3$Stellar Astrophysics Centre, Department of Physics and Astronomy, Aarhus University, Ny Munkegade 120, DK-8000 Aarhus C, Denmark\\
$^4$LESIA, Observatoire de Paris, PSL Research University, CNRS, Sorbonne Universit\'{e}s, UPMC Univ. Paris 06, Univ. Paris Diderot, \\
~~Sorbonne Paris Cit\'{e}, 5 place Jules Janssen, 92195 Meudon, France
}
\date{Accepted XXX. Received YYY; in original form ZZZ}

\pubyear{2019}

\usepackage{etoolbox}
\makeatletter
\patchcmd\@combinedblfloats{\box\@outputbox}{\unvbox\@outputbox}{}{%
   \errmessage{\noexpand\@combinedblfloats could not be patched}%
}%
 \makeatother

\begin{document}
\label{firstpage}
\pagerange{\pageref{firstpage}--\pageref{lastpage}}
\maketitle

\begin{abstract}
We report our survey of \gdor stars from the 4-yr \textit{Kepler} mission. These stars pulsate mainly in g modes and r modes, showing period-spacing patterns in the amplitude spectra. The period-spacing patterns are sensitive to the chemical composition gradients and the near-core rotation, hence they are essential for understanding the stellar interior.
We identified period-spacing patterns in \starnumber \gdor stars. Almost every star pulsates in dipole g modes, while about 30\% of stars also show clear patterns for quadrupole g modes and 16\% of stars present r mode patterns. We measure periods, period spacings, and the gradient of the period spacings. These three observables guide the mode identifications and can be used to estimate the near-core rotation rate. We find many stars are hotter and show longer period-spacing patterns than theory. 
Using the Traditional Approximation of Rotation (TAR), we inferred the asymptotic spacings, the near-core rotation rates, and the radial orders of the g and r modes. 
Most stars have a near-core rotation rate around $1$\,\cpd and an asymptotic spacing around 4000\,s. We also find that many stars rotate more slowly than predicted by theory for unclear reasons. \numberfastsplitting stars show rotational splittings with fast rotation rates. We compared the observed slope--rotation relation with the theory and find a large spread. We detected rotational modulations in \numbersurfacemodulation stars and used them to derive the core-to-surface rotation ratios. The interiors rotate faster than the cores in most stars, but by no more than 5\%. 
\end{abstract}



\section{Introduction}
Rotation affects the transport of chemical elements and angular momentum in stars, so it changes stellar structure and evolution \citep[e.g.][]{Maeder_2009, Mathis_2013}. However, the theoretical description of rotation is still a matter of debate. For example, the observed core-to-surface rotation rate ratios in red giants are smaller than predicted by theory \citep[see e.g.][]{Mosser_2012, Eggenberger_2012, Ceillier_2013,Fuller_2019}. 
For A- and F-type main-sequence stars, the typical value of the projected surface rotation velocity is around $100\,\mathrm{km\,s^{-1}}$ and increases with effective temperature \citep[e.g.][]{Fukuda_1982, Groot_1996, Abt_1995, Royer_2007}. Hence, the effect of rapid rotation must be treated properly. 

Stellar oscillations are a powerful tool to investigate the stellar interior. We focus on $\gamma$\,Doradus stars, which are A- to F-type main-sequence stars with typical masses from 1.4 to 2.0\,$\mathrm{M_\odot}$ \citep[e.g.][]{Kaye_1999, VanReeth_2016_TAR}. The pulsations of $\gamma$\,Dor stars are gravity modes with high radial order ($20 \lesssim n  \lesssim 100$) low degree ($l\lesssim 4$) with typical pulsation period from 0.3 to 3\,d \citep{Balona_1994, Kaye_1999, Saio_2018, VanReeth_2018, Li_2019}. 
\textcolor{black}{Gravity modes have their highest mode energy in the near-core regions \citep[e.g.][]{Triana_2015, VanReeth_2016_TAR}. Therefore, \gdor stars allow us to investigate the stellar interior. 
The excitation mechanism of \gdor stars is still in debate. \cite{Guzik_2000, Dupret_2005} reported that the g-mode pulsations are excited by the convective flux blocking mechanism that operates at the base of the envelope convection zone. \cite{Xiong_2016} found that the radiative $\kappa$ mechanism plays a major role in warm \gdor stars while the coupling between convection and oscillations is dominant in cool stars. Turbulent thermal convection is a damping mechanism that gives rise to the red edge of the instability strip. \cite{Grassitelli_2015} pointed out that turbulent pressure fluctuations may contribute to the \gdor phenomenon.} 
The instability strip of \gdor stars is located between the solar-like stars and the $\delta$\,Scuti stars, overlapping with the red edge of the $\delta$\,Scuti instability strip \citep{Dupret_2005, Bouabid_2009, Bouabid_2013}. Hence some \gdor stars show both g- and p-mode oscillations and are called $\delta$\,Sct--\gdor hybrids. Pressure modes probe the outer stellar layers, therefore the overall structure along the radial direction can be deduced \citep[e.g.][]{Kurtz_2014, Saio_2015}.

Due to the daily aliasing and small amplitudes, the pulsations of \gdor stars were hard to detect with ground-based observations, hence their near-core rotations were unclear for a long time. Thanks to the \textit{Kepler} space telescope \citep{Koch_2010, Borucki_2010}, 4-yr light curves of many stars have been collected. \cite{Kurtz_2014} measured the rotational splittings of the \gdor star KIC\,11145123, which was the first robust determination of the rotation of the deep core and surface of a main-sequence star. The rotational splittings of g modes in Slowly Pulsating B (SPB) stars were also reported \cite[e.g.][]{Papics_2015}. For fast rotators, as in the majority of \gdor stars, the period spacings change quasi-linearly with period and can be used to fit the near-core rotation rate \citep[e.g.][]{VanReeth_2016_TAR, Christophe_2018}. Now, splittings or period spacings of g modes from tens of \gdor stars were found, both in single stars and binaries, whose rotation profiles are almost uniform \citep[e.g.][]{Triana_2015, Saio_2015, Keen_2015, VanReeth_2015, Li_2018, Guo2017, Kallinger_2017, Li_2019}. 

In this paper, we report \patternnumber period-spacing patterns from \starnumber \textit{Kepler} \gdor stars, which form the largest sample of identified period-spacing patterns. The period spacing $\Delta P$ is defined as the period difference between two consecutive overtones $\Delta P_n \equiv P_{n+1}-P_{n}$ and is a constant in the non-rotating homogeneous stars \citep[according to the asymptotic relation,][]{Shibahashi1979}. The rapid rotation changes stellar structure and frequency values, and the Traditional Approximation of Rotation (TAR), or the complete calculation including the full effect of rotation is necessary to describe the oscillation frequencies more accurately \citep[e.g.][]{Eckart_1960, Lee1987, Townsend2005, Saio_2018}. Under the TAR, the period spacing decreases with period quasi-linearly for the prograde and zonal g modes. Overall, the retrograde g modes have increasing period spacing \citep{Bouabid_2013, Ouazzani_2017} and they are seen in slow rotators but are hard to see in fast rotators \citep{Li_2018, Saio_2018}.

In addition to g modes and sometimes p modes, \gdor stars also show Rossby modes (r modes), whose restoring force is the Coriolis force \citep{Papaloizou_1978_first_r_mode}. Rossby modes propagate retrograde to the rotation direction and have discrete frequencies smaller than the rotation frequency in the corotating reference frame \citep{Saio_1982_Rossby_mode, Lee_1997_rotating, Provost_1981_eigenfreq_of_Rossby_mode}. Rossby modes can also be described by the TAR and they also show a quasi-linear period-spacing pattern, in which the period spacing increases with period. 
Using the period-spacing patterns from g and r modes, the near-core rotation of tens of \gdor stars were measured to be around $1\,\mathrm{d^{-1}}$ \citep[e.g.][]{Saio_2018_Rossby_mode, VanReeth_2018, Li_2019}. Many of the stars in our sample also show r modes.

We describe our data reduction and TAR fitting in Section~\ref{sec:data_reduction}. Section~\ref{sec:results} gives the observation results, including the observed relative occurence rates of different types of modes, the typical structure of the periodogram, the slope--mean period relation of \gdor stars. Section~\ref{sec:TAR_results} shows the TAR fit results, including the distributions of the asymptotic spacings and the near-core rotation rates, as well as the comparison with the theoretical predictions. Section~\ref{sec:rot_on_S_P_diagram} reveals that the slope--period relation can be used on estimating the near-core rotation rate. Section~\ref{sec:fast_splitting} reports \numberfastsplitting fast-rotating stars with rotational splittings and Section~\ref{sec:surface_rotation} displays \numbersurfacemodulation stars with surface rotation modulations. Finally, we conclude our works in Section~\ref{sec:conclusions}.

\section{Data analysis}\label{sec:data_reduction}

\begin{figure*}
    \centering
    \includegraphics[width=1\linewidth]{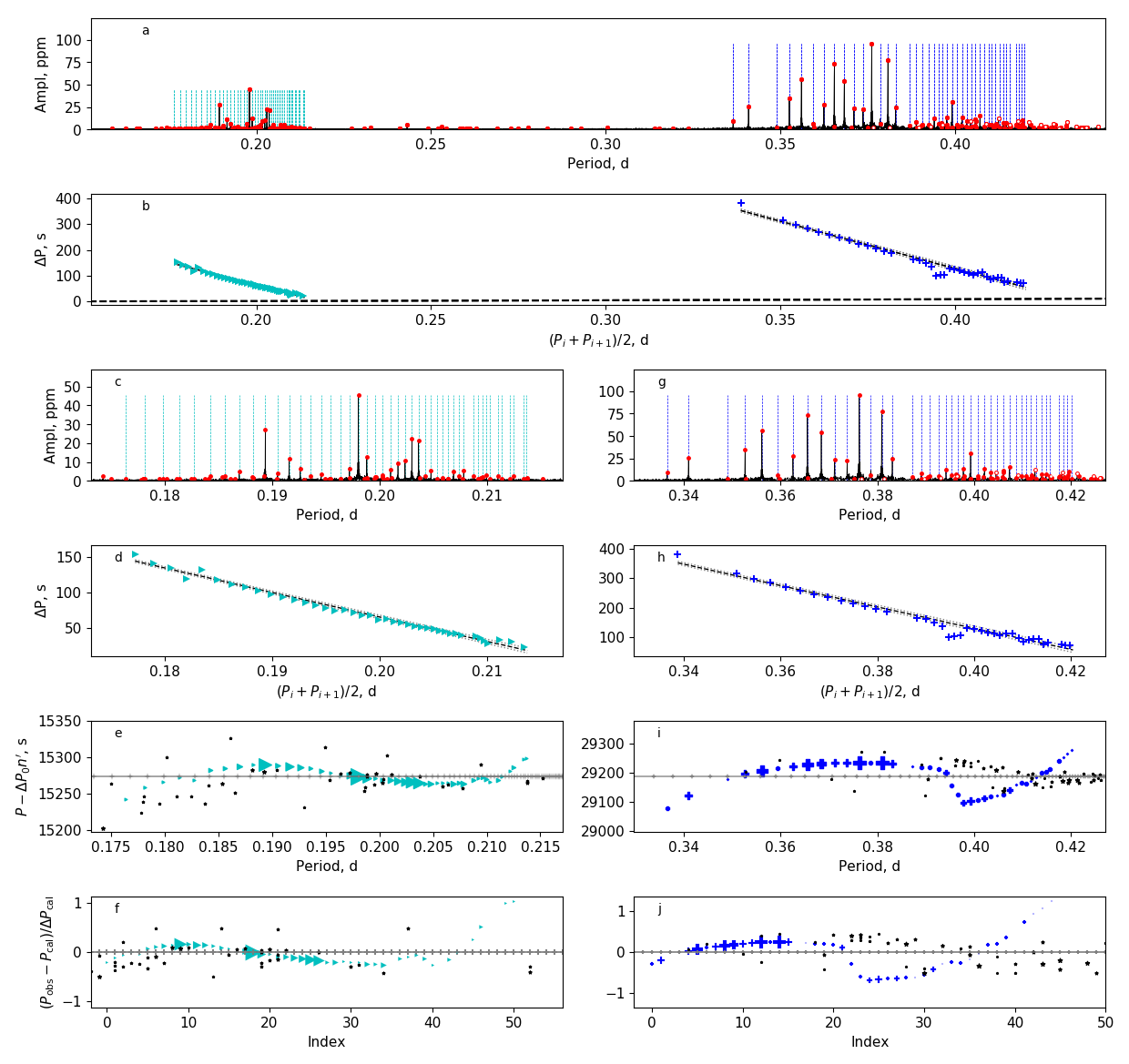}
    \caption{The g-mode patterns of KIC\,7694191. Panel a: the amplitude spectrum with x-axis of period. The y-axis is the amplitude in ppm. The solid red circles present the detected independent frequencies while the open red circles show the combination frequencies. The vertical dashed lines are the linear fits for each pattern. We found two independent frequency groups around 0.20\,d and 0.38\,d. There are two period-spacing patterns. The blue one on the right is the $l=1,~m=1$ g modes while the cyan one on the left is the $l=2,~m=2$ g modes, whose periods are marked by the vertical dashed lines. Panel b: the period-spacing patterns of KIC\,7694191. The linear fits and uncertainties are shown by the black and grey dashed lines with dips removed. The blue plus symbols are the dipole g modes and the cyan triangle symbols are the quadrupole g modes. Panels c and d: the detail of the spectrum and period-spacing pattern of quadrupole g modes. Panel e: the sideways \'{e}chelle diagram of the quadrupole g-modes pattern. The cyan triangles are the periods belonging to the pattern while the black stars are the noisy peaks. Panel f: the normalised sideways \'{e}chelle diagram of the quadrupole g-modes pattern. Panels g to j: same as (c -- f) but for the dipole g-mode patterns. }
    \label{fig:KIC7694191}
\end{figure*}

\begin{figure}
    \centering
    \includegraphics[width=1\linewidth]{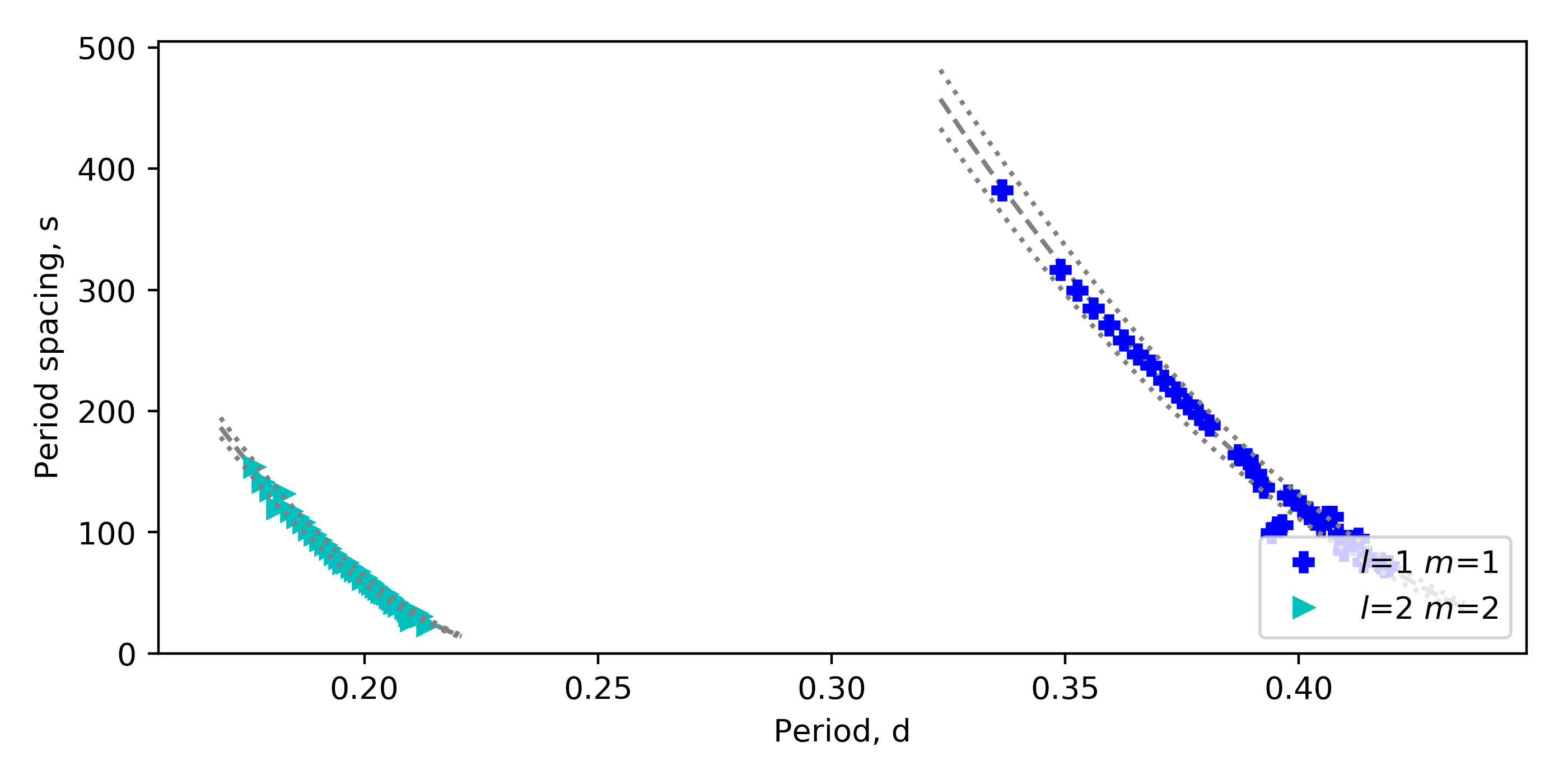}
    \caption{The TAR fitting of KIC\,7694191. The dashed lines are the best-fitting result and the dotted lines show the uncertainty. }
    \label{fig:KIC7694191_TAR_fitting}
\end{figure}

We examined \totalnumber stars with effective temperature between 6000\,K to 10000\,K, where we used the input temperatures from the \textit{Kepler} DR25 data release \citep{Mathur_2017}. We found \patternnumber clear period-spacing patterns in \starnumber stars, including 50 stars by \cite{VanReeth_2015}, 22 stars with splittings by \cite{Li_2018}, 82 stars with r-mode patterns by \cite{Li_2019}, 30 stars by \cite{Chowdhury_2018}, 44 stars by \cite{Murphy_2018}, 344 stars found by Barbara et al. (in prep), and the rest we found by visually inspecting light curves and their Fourier transforms (the samples overlap).
Barbara et al. (in prep) applied a Gaussian mixture model in a reduced 5D space to classify 12066 stars in the \textit{Kepler} field. The method involves using a greedy algorithm to select defining features from the HCTSA feature library \citep{fulcher_2013, fulcher_2017}.

We used 4-year \textit{Kepler} long-cadence light curves (29.45-min sampling) from the multi-scale MAP data pipeline \citep{Stumpe_2014}. \textcolor{black}{The 4-yr long-cadence data are suitable for \gdor stars since the typical pulsation periods of these stars are around 1 day with period spacings around 1000\,s, which require a long observation span to resolve the modes. However, \shortcadencedatanumber of our stars also have short-cadence data. These data can be used to readily investigate the pressure modes if they are $\gamma$\,Dor--$\delta$\,Sct hybrids, though thanks to the 
super-Nyquist asteroseismology technique \citep{Murphy_2013} the \textit{Kepler} LC data are also sufficient for this purpose. }

In each quarter, the light curve was divided by a second-order polynomial fit to remove any slow trend. We computed the Fourier transform and extracted the frequencies until the signal to noise ratio (S/N) was smaller than 3. 

The period-spacing patterns were identified by the cross-correlation algorithm described by \cite{Li_2018} and inspected visually. We present the period-spacing patterns of KIC\,7694191 as an example in Fig.~\ref{fig:KIC7694191}. Figure~\ref{fig:KIC7694191}a shows the periodogram, where the locations of peaks for each pattern are shown with dashed vertical lines. We found two period-spacing patterns around 0.20\,d and 0.38\,d. The right pattern in Fig.~\ref{fig:KIC7694191}a comprises the dipole ($l=1$) sectoral ($m=l=1$) g modes while the left one comprises the quadrupole ($l=2$) sectoral ($m=l=2$) g modes. 
The mode identifications were based on the TAR fit and \cite{Saio_2018}, as described below.

Fig.~\ref{fig:KIC7694191}b presents the period spacing versus period. The period spacing for the dipole g modes decreases from 400\,s to 100\,s with increasing period. For the quadrupole g-mode pattern, the period spacing drops from 150\,s to 50\,s with increasing period. Both patterns show deviations from the linear model, such as the dip at 0.39\,d in the dipole g-mode pattern. In a rapidly-rotating star, the dip is more likely to form because of the mode coupling between sectoral and tesseral modes \citep{Saio_2018}. The linear fits and their uncertainties with the dips removed are shown by the black and grey dashed lines. Hence the linear fits are not affected by the dips. 

After obtaining the initial estimates for the parameters from the cross-correlation algorithm, the sideways \'{e}chelle diagram was made based on the formula
\begin{equation}
    P_i=\Delta P_0 \frac{\left(1+\Sigma\right)^i-1}{\Sigma}+P_0=\Delta P_0\left(n'+\epsilon \right),\label{equ:P_i}
\end{equation}
with the assumption that the period spacing changes linearly with period. Here, $P_i$ is the $i^\mathrm{th}$ pulsation period, $P_0$ is the first pulsation period, $\Delta P_0$ is the first period spacing, $\Sigma$ is the slope in the linear assumption, $n'\equiv \frac{\left(1+\Sigma\right)^i-1}{\Sigma}$ is the normalised index, and $\epsilon$ is the ratio $P_0/\Delta P_0$ \citep{Li_2018}. The x-axis of the sideways \'{e}chelle diagram is the observed period and the y-axis is the difference between the observed and fitted periods from Eq.~\ref{equ:P_i}. 

Panels (c) and (d) zoom in on the quadrupole g modes from panels (a) and (b), while panels (g) and (h) do the same for the dipole g modes. In Panels (e) and (i), the \'{e}chelle diagrams are plotted sideways. The x-axis is the pulsation period while the y-axis is the term $P^{\mathrm{obs}}-n'\Delta P$ from the fit of eq.~\ref{equ:P_i}. During this fit, we did not exclude any dips. For the peaks that do not belong to the pattern, we plotted them at the location that minimised the value $P^{\mathrm{obs}}-n'\Delta P$. Therefore, the y-axis reflects the deviation from the linear fit, similar to the curvature in the \'{e}chelle diagram of solar-like oscillators \citep[e.g.][]{Mazumdar_2014}. In panel (e), the curve is smooth and is dominated by the slightly changing slope in the quadrupole sectoral g modes. In panel (i), there is a rapid drop at 0.39\,d, which is caused by the dip here. Panels (f) and (j) show the normalised sideways \'{e}chelle diagram. The x-axis is the index of peaks, counting the first peak as 0, and the y-axis is the deviation over the local period spacing $\left(P^{\mathrm{obs}}-n'\Delta P\right)/\Delta P$ expressed as a percentage.

For each period-spacing pattern, consisting of a series of pulsation periods $\lbrace P_i \rbrace$, we measured three observables: the mean period, the mean period spacing, and the slope. The mean period $\langle P \rangle$ is the average of the pulsation periods. The mean period spacing $\langle \Delta P \rangle$ is the slope of the linear fit between the periods $P_i$ and the index $i$. The slope $\Sigma$ is the changing rate between the period spacing and the period with dips removed.

After identifying a period-spacing pattern, the asymptotic formulation of the Traditional Approximation of Rotation (TAR) was used to fit the pattern assuming rigid rotation \citep[e.g.][]{Eckart_1960, Lee_1997_rotating, Townsend_2003, VanReeth_2016_TAR}. The pulsation periods in the co-rotating reference frame were computed by
\begin{equation}
P^\mathrm{TAR}_{nlm, \mathrm{co}}=\frac{\Pi_0}{\sqrt{\lambda_{l,m,s}}}\left(n+ \varepsilon_g \right),  \label{equ:TAR_P}
\end{equation}
where $\Pi_0$ is the asymptotic period spacing, $n$ is the radial order, the phase term $\varepsilon_g$ is set as $0.5$, and $\lambda_{l,m,s}$ is the eigenvalue of the Laplace tidal equation, which is specified by the angular degree $l$ for g modes or the value $k$ for r modes, the azimuthal order $m$ and the spin parameter~$s$. The value $k$ is used since the angular degree $l$ is not defined for r modes \citep{Lee_1997_rotating}. The spin parameter is defined as
\begin{equation}
s\equiv\frac{2f_\mathrm{rot}}{f_\mathrm{co}},\label{equ:spin_parameter}
\end{equation}
where $f_\mathrm{rot}$ is the rotation frequency and $f_\mathrm{co}$ is the pulsation frequency in the co-rotating frame. The TAR periods in the inertial reference frame are given by 
\begin{equation}
   P^\mathrm{TAR}_{nlm, \mathrm{in}} =\frac{1}{1/P^\mathrm{TAR}_{nlm, \mathrm{co}}+m f_\mathrm{rot}}.
\end{equation}
Hence the near-core rotation rate, the asymptotic spacing, and the radial orders can be obtained by fitting these pulsation periods to the observed pattern using a Markov chain Monte Carlo (MCMC) optimising code described by \cite{Li_2019}. Fig.~\ref{fig:KIC7694191_TAR_fitting} presents the TAR fitting result of KIC\,7694191. The near-core rotation rate is $2.083\pm0.009$\,\cpd and the asymptotic spacing is $4400\pm200\,\mathrm{s}$. The best-fitting curves (dashed lines) follow the observed pattern and show slowly-changing slopes with period.

\section{Results}\label{sec:results}

The parameters of the stars, the period-spacing patterns and the TAR fit results are listed in the online-only table, while table~\ref{tab:rot_Pi0_table} shows part of the table for guidance on style and content. \textcolor{black}{We also indicate which of the \shortcadencedatanumber stars have short-cadence data and we indicated which of the \hybridsnumber stars show significant pressure modes oscillations. \hybridsshortcadencenumber ~have both. These give the possibility to investigate core-to-surface physics by using g and p modes together.} All the period-spacing patterns are shown in Appendix~\ref{appendix:period_spacing_patterns}, which is also online-only.

\begin{landscape}
\begin{table} 
\centering
\caption{\textcolor{black}{KIC numbers, Kepler magnitudes $K_p$, effective temperatures $T_\mathrm{eff}$, luminosities $L$, mode identifications (for g modes, we give the angular degrees $l$ and the azimuthal orders $m$, While for r modes, we list the value $k$ and $m$), mean pulsation periods $\langle P \rangle$, mean period spacings $\langle \Delta P \rangle$, slopes $\Sigma$, asymptotic spacings $\Pi_0$, near-core rotation rates $f_\mathrm{rot}$, the ranges of radial orders $n$, and ranges of spin parameters $s$ of \starnumber stars in this paper. For the last two columns, `1' marks the stars which have short-cadence data (SC) or are $\gamma$\,Dor--$\delta$\,Sct hybrids (H).}}\label{tab:rot_Pi0_table} 
\begin{tabular}{rrrrrrrrrrrrrrrrrr} 
\hline
KIC & $K_p$ & $T_\mathrm{eff}$ & $L$ & $l$ & $k$ & $m$ & $\langle P \rangle$ & $\langle \Delta P \rangle$ & $\Sigma$ & $\Pi_0$ & $f_\mathrm{rot}$ & \multicolumn{2}{c}{$n$} & \multicolumn{2}{c}{$s$} & SC & H \\ 
   &   & K & $\mathrm{L_\odot}$ &  &  &  &   days   &    Seconds      &   days/days   & Seconds & $\mathrm{d^{-1}} $ &   min   &  max   & min          &     max  \\ 
\hline
1026861 & $11.00$ & $7060\pm80$\phantom{$0$} & $8.5\pm0.5$ & $2$ &       & \phantom{$-$}$2$ & $0.8$ & $530$ & \phantom{$-$}$-0.0123 \pm 0.0002$\phantom{$0$} & \phantom{$-$}$4310\pm 50$\phantom{$0$}\phantom{$0$} & \phantom{$-$}$0.311\pm 0.002$\phantom{$0$} & $\phantom{0}57$ & $\phantom{0}92$ & $\phantom{0}0.7$ & $\phantom{0}1.3$ & $1$ & \phantom{$1$} \\
        &         &         &         & $1$ &       & \phantom{$-$}$1$ & $1.5$ & $1240$ & \phantom{$-$}$-0.0167 \pm 0.0007$\phantom{$0$} &         &         & $\phantom{0}45$ & $\phantom{0}79$ & $\phantom{0}1.1$ & $\phantom{0}2.1$ & \phantom{$1$} & \phantom{$1$} \\
1160891 & $13.21$ & $6840\pm80$\phantom{$0$} &        & $1$ &       & \phantom{$-$}$1$ & $0.6$ & $200$ & \phantom{$-$}$-0.022 \pm 0.002$\phantom{$0$}\phantom{$0$} & \phantom{$-$}$2800\pm 400$\phantom{$0$} & \phantom{$-$}$1.14\pm 0.03$\phantom{$0$}\phantom{$0$} & $\phantom{0}66$ & $\phantom{0}88$ & $\phantom{0}4.5$ & $\phantom{0}6.2$ & \phantom{$1$} & \phantom{$1$} \\
1162345 & $11.68$ & $6500\pm200$ & $38\pm3$\phantom{$.0$} & $2$ &       & \phantom{$-$}$2$ & $0.3$ & $70$ & \phantom{$-$}$-0.0275 \pm 0.0002$\phantom{$0$} & \phantom{$-$}$4400\pm 400$\phantom{$0$} & \phantom{$-$}$1.46\pm 0.02$\phantom{$0$}\phantom{$0$} & $\phantom{0}55$ & $\phantom{0}71$ & $\phantom{0}3.9$ & $\phantom{0}5.1$ & \phantom{$1$} & \phantom{$1$} \\
        &         &         &         & $1$ &       & \phantom{$-$}$1$ & $0.5$ & $290$ & \phantom{$-$}$-0.0372 \pm 0.0003$\phantom{$0$} &         &         & $\phantom{0}35$ & $\phantom{0}47$ & $\phantom{0}4.8$ & $\phantom{0}6.6$ & \phantom{$1$} & \phantom{$1$} \\
1295531 & $11.94$ & $6760\pm80$\phantom{$0$} &        & $1$ &       & \phantom{$-$}$1$ & $0.8$ & $1280$ & \phantom{$-$}$-0.0266 \pm 0.0002$\phantom{$0$} & \phantom{$-$}$4200\pm 20$\phantom{$0$}\phantom{$0$} & \phantom{$-$}$0.572\pm 0.002$\phantom{$0$} & $\phantom{0}18$ & $\phantom{0}50$ & $\phantom{0}0.8$ & $\phantom{0}2.5$ & \phantom{$1$} & \phantom{$1$} \\
1431379 & $12.62$ & $6670\pm80$\phantom{$0$} & $8.5\pm0.7$ & $2$ &       & \phantom{$-$}$2$ & $0.3$ & $200$ & \phantom{$-$}$-0.0384 \pm 0.0006$\phantom{$0$} & \phantom{$-$}$4360\pm 10$\phantom{$0$}\phantom{$0$} & \phantom{$-$}$1.2526\pm 0.0006$ & $\phantom{0}21$ & $\phantom{0}59$ & $\phantom{0}1.2$ & $\phantom{0}3.6$ & \phantom{$1$} & \phantom{$1$} \\
        &         &         &         & $1$ &       & \phantom{$-$}$1$ & $0.6$ & $330$ & \phantom{$-$}$-0.0349 \pm 0.0005$\phantom{$0$} &         &         & $\phantom{0}25$ & $\phantom{0}66$ & $\phantom{0}2.9$ & $\phantom{0}8.1$ & \phantom{$1$} & \phantom{$1$} \\
        &         &         &         &       & $-2$ & $-1$ & $1.0$ & $390$ & $0.0504 \pm 0.0004$\phantom{$0$} &         &         & $\phantom{0}12$ & $\phantom{0}60$ & $\phantom{0}7.1$ & $23.1$ & \phantom{$1$} & \phantom{$1$} \\
1432149 & $11.22$ & $7500\pm300$ & $11.6\pm0.8$ & $2$ &       & \phantom{$-$}$2$ & $0.3$ & $80$ & \phantom{$-$}$-0.0269 \pm 0.0003$\phantom{$0$} & \phantom{$-$}$4200\pm 400$\phantom{$0$} & \phantom{$-$}$1.38\pm 0.02$\phantom{$0$}\phantom{$0$} & $\phantom{0}54$ & $\phantom{0}75$ & $\phantom{0}3.4$ & $\phantom{0}4.8$ & $1$ & \phantom{$1$} \\
        &         &         &         & $1$ &       & \phantom{$-$}$1$ & $0.5$ & $260$ & \phantom{$-$}$-0.0320 \pm 0.0004$\phantom{$0$} &         &         & $\phantom{0}44$ & $\phantom{0}52$ & $\phantom{0}5.5$ & $\phantom{0}6.5$ & \phantom{$1$} & \phantom{$1$} \\
1575977 & $13.61$ & $7300\pm300$ & $8.8\pm0.7$ & $1$ &       & \phantom{$-$}$1$ & $0.4$ & $710$ & \phantom{$-$}$-0.0543 \pm 0.0007$\phantom{$0$} & \phantom{$-$}$3800\pm 100$\phantom{$0$} & \phantom{$-$}$1.50\pm 0.02$\phantom{$0$}\phantom{$0$} & $\phantom{0}18$ & $\phantom{0}28$ & $\phantom{0}2.0$ & $\phantom{0}3.3$ & \phantom{$1$} & $1$ \\
1872262 & $13.74$ & $7100\pm200$ & $7.6\pm0.6$ & $2$ &       & \phantom{$-$}$2$ & $0.5$ & $260$ & \phantom{$-$}$-0.028 \pm 0.002$\phantom{$0$}\phantom{$0$} & \phantom{$-$}$5000\pm 2000$ & \phantom{$-$}$0.70\pm 0.05$\phantom{$0$}\phantom{$0$} & $\phantom{0}50$ & $\phantom{0}58$ & $\phantom{0}2.0$ & $\phantom{0}2.4$ & \phantom{$1$} & \phantom{$1$} \\
1996456 & $11.44$ & $7100\pm200$ & $10.1\pm0.8$ & $2$ &       & \phantom{$-$}$2$ & $0.4$ & $160$ & \phantom{$-$}$-0.0294 \pm 0.0002$\phantom{$0$} & \phantom{$-$}$4450\pm 80$\phantom{$0$}\phantom{$0$} & \phantom{$-$}$1.038\pm 0.004$\phantom{$0$} & $\phantom{0}30$ & $\phantom{0}84$ & $\phantom{0}1.4$ & $\phantom{0}4.3$ & $1$ & \phantom{$1$} \\
        &         &         &         & $1$ &       & \phantom{$-$}$1$ & $0.7$ & $420$ & \phantom{$-$}$-0.0338 \pm 0.0002$\phantom{$0$} &         &         & $\phantom{0}30$ & $\phantom{0}63$ & $\phantom{0}2.8$ & $\phantom{0}6.3$ & \phantom{$1$} & \phantom{$1$} \\
2018685 & $14.05$ & $6700\pm200$ & $8.4\pm0.7$ & $2$ &       & \phantom{$-$}$2$ & $0.3$ & $160$ & \phantom{$-$}$-0.0315 \pm 0.0004$\phantom{$0$} & \phantom{$-$}$4200\pm 200$\phantom{$0$} & \phantom{$-$}$1.27\pm 0.01$\phantom{$0$}\phantom{$0$} & $\phantom{0}32$ & $\phantom{0}57$ & $\phantom{0}1.8$ & $\phantom{0}3.4$ & \phantom{$1$} & \phantom{$1$} \\
        &         &         &         & $1$ &       & \phantom{$-$}$1$ & $0.6$ & $350$ & \phantom{$-$}$-0.0343 \pm 0.0005$\phantom{$0$} &         &         & $\phantom{0}35$ & $\phantom{0}50$ & $\phantom{0}4.1$ & $\phantom{0}5.9$ & \phantom{$1$} & \phantom{$1$} \\
2020444 & $13.00$ & $6890\pm80$\phantom{$0$} & $8.9\pm0.7$ & $1$ &       & \phantom{$-$}$1$ & $1.0$ & $1040$ & \phantom{$-$}$-0.0211 \pm 0.0004$\phantom{$0$} & \phantom{$-$}$4360\pm 30$\phantom{$0$}\phantom{$0$} & \phantom{$-$}$0.499\pm 0.002$\phantom{$0$} & $\phantom{0}30$ & $\phantom{0}66$ & $\phantom{0}1.2$ & $\phantom{0}3.1$ & \phantom{$1$} & \phantom{$1$} \\
2141387 & $12.15$ & $7200\pm300$ & $7.0\pm0.6$ & $1$ &       & \phantom{$-$}$1$ & $1.0$ & $1010$ & \phantom{$-$}$-0.035 \pm 0.001$\phantom{$0$}\phantom{$0$} & \phantom{$-$}$6090\pm 90$\phantom{$0$}\phantom{$0$} & \phantom{$-$}$0.592\pm 0.003$\phantom{$0$} & $\phantom{0}29$ & $\phantom{0}51$ & $\phantom{0}2.1$ & $\phantom{0}4.0$ & \phantom{$1$} & \phantom{$1$} \\
2163896 & $13.12$ & $6900\pm200$ & $4.8\pm0.3$ & $1$ &       & \phantom{$-$}$1$ & $0.8$ & $460$ & \phantom{$-$}$-0.0249 \pm 0.0002$\phantom{$0$} & \phantom{$-$}$3400\pm 100$\phantom{$0$} & \phantom{$-$}$0.834\pm 0.008$\phantom{$0$} & $\phantom{0}44$ & $\phantom{0}74$ & $\phantom{0}2.5$ & $\phantom{0}4.5$ & \phantom{$1$} & \phantom{$1$} \\
2168333 & $10.08$ & $8400\pm300$ & $34\pm3$\phantom{$.0$} & $1$ &       & \phantom{$-$}$1$ & $0.5$ & $220$ & \phantom{$-$}$-0.0358 \pm 0.0004$\phantom{$0$} & \phantom{$-$}$4200\pm 200$\phantom{$0$} & \phantom{$-$}$1.66\pm 0.01$\phantom{$0$}\phantom{$0$} & $\phantom{0}31$ & $\phantom{0}62$ & $\phantom{0}4.5$ & $\phantom{0}9.5$ & $1$ & $1$ \\
2300165 & $11.05$ & $7400\pm300$ & $7.2\pm0.6$ & $2$ &       & \phantom{$-$}$2$ & $0.3$ & $170$ & \phantom{$-$}$-0.0300 \pm 0.0002$\phantom{$0$} & \phantom{$-$}$4140\pm 70$\phantom{$0$}\phantom{$0$} & \phantom{$-$}$1.136\pm 0.004$\phantom{$0$} & $\phantom{0}29$ & $\phantom{0}75$ & $\phantom{0}1.4$ & $\phantom{0}3.9$ & $1$ & \phantom{$1$} \\
        &         &         &         & $1$ &       & \phantom{$-$}$1$ & $0.6$ & $380$ & \phantom{$-$}$-0.0323 \pm 0.0003$\phantom{$0$} &         &         & $\phantom{0}26$ & $\phantom{0}71$ & $\phantom{0}2.4$ & $\phantom{0}7.4$ & \phantom{$1$} & \phantom{$1$} \\
2309579 & $13.35$ & $7200\pm300$ & $10.0\pm0.7$ & $2$ &       & \phantom{$-$}$2$ & $0.4$ & $110$ & \phantom{$-$}$-0.0211 \pm 0.0002$\phantom{$0$} & \phantom{$-$}$4000\pm 200$\phantom{$0$} & \phantom{$-$}$1.001\pm 0.007$\phantom{$0$} & $\phantom{0}58$ & $\phantom{0}95$ & $\phantom{0}2.5$ & $\phantom{0}4.2$ & \phantom{$1$} & \phantom{$1$} \\
        &         &         &         & $1$ &       & \phantom{$-$}$1$ & $0.8$ & $210$ & \phantom{$-$}$-0.0230 \pm 0.0001$\phantom{$0$} &         &         & $\phantom{0}50$ & $107$ & $\phantom{0}4.3$ & $\phantom{0}9.4$ & \phantom{$1$} & \phantom{$1$} \\
2449383 & $13.92$ & $7200\pm300$ & $7.6\pm0.6$ & $2$ &       & \phantom{$-$}$2$ & $0.2$ & $90$ & \phantom{$-$}$-0.0310 \pm 0.0003$\phantom{$0$} & \phantom{$-$}$3910\pm 90$\phantom{$0$}\phantom{$0$} & \phantom{$-$}$1.588\pm 0.007$\phantom{$0$} & $\phantom{0}35$ & $\phantom{0}76$ & $\phantom{0}2.3$ & $\phantom{0}5.2$ & \phantom{$1$} & \phantom{$1$} \\
        &         &         &         & $1$ &       & \phantom{$-$}$1$ & $0.4$ & $320$ & \phantom{$-$}$-0.0419 \pm 0.0004$\phantom{$0$} &         &         & $\phantom{0}20$ & $\phantom{0}59$ & $\phantom{0}2.5$ & $\phantom{0}8.0$ & \phantom{$1$} & \phantom{$1$} \\
2450944 & $15.74$ & $6800\pm200$ &        & $1$ &       & \phantom{$-$}$1$ & $1.0$ & $2500$ & \phantom{$-$}$-0.0084 \pm 0.0004$\phantom{$0$} & \phantom{$-$}$4136\pm 5$\phantom{$0$}\phantom{$0$}\phantom{$0$} & \phantom{$-$}$0.141\pm 0.001$\phantom{$0$} & $\phantom{0}24$ & $\phantom{0}37$ & $\phantom{0}0.2$ & $\phantom{0}0.3$ & \phantom{$1$} & $1$ \\
        &         &         &         & $1$ &       & $-1$ & $1.2$ & $3350$ & $0.003 \pm 0.001$\phantom{$0$}\phantom{$0$} &         &         & $\phantom{0}30$ & $\phantom{0}37$ & $\phantom{0}0.2$ & $\phantom{0}0.3$ & \phantom{$1$} & \phantom{$1$} \\
2575161 & $10.88$ & $6900\pm100$ &        & $1$ &       & \phantom{$-$}$1$ & $0.4$ & $250$ & \phantom{$-$}$-0.0466 \pm 0.0008$\phantom{$0$} & \phantom{$-$}$4470\pm 20$\phantom{$0$}\phantom{$0$} & \phantom{$-$}$1.833\pm 0.001$\phantom{$0$} & $\phantom{0}20$ & $\phantom{0}54$ & $\phantom{0}3.6$ & $10.0$ & \phantom{$1$} & \phantom{$1$} \\
        &         &         &         &       & $-2$ & $-1$ & $0.7$ & $580$ & $0.0753 \pm 0.0004$\phantom{$0$} &         &         & $\phantom{0}\phantom{0}8$ & $\phantom{0}27$ & $\phantom{0}7.3$ & $16.2$ & \phantom{$1$} & \phantom{$1$} \\
2578582 & $13.93$ & $7300\pm300$ & $5.2\pm0.4$ & $1$ &       & \phantom{$-$}$1$ & $0.8$ & $1140$ & \phantom{$-$}$-0.0233 \pm 0.0003$\phantom{$0$} & \phantom{$-$}$3960\pm 30$\phantom{$0$}\phantom{$0$} & \phantom{$-$}$0.570\pm 0.003$\phantom{$0$} & $\phantom{0}24$ & $\phantom{0}53$ & $\phantom{0}1.0$ & $\phantom{0}2.5$ & \phantom{$1$} & \phantom{$1$} \\
2579147 & $13.83$ & $7300\pm300$ & $14\pm1$\phantom{$.0$} & $2$ &       & \phantom{$-$}$2$ & $0.7$ & $620$ & \phantom{$-$}$-0.0096 \pm 0.0004$\phantom{$0$} & \phantom{$-$}$3600\pm 30$\phantom{$0$}\phantom{$0$} & \phantom{$-$}$0.286\pm 0.003$\phantom{$0$} & $\phantom{0}54$ & $\phantom{0}72$ & $\phantom{0}0.5$ & $\phantom{0}0.7$ & \phantom{$1$} & $1$ \\
        &         &         &         & $1$ &       & \phantom{$-$}$1$ & $1.1$ & $1560$ & \phantom{$-$}$-0.0154 \pm 0.0006$\phantom{$0$} &         &         & $\phantom{0}43$ & $\phantom{0}51$ & $\phantom{0}0.8$ & $\phantom{0}0.9$ & \phantom{$1$} & \phantom{$1$} \\
2696217 & $13.45$ & $7400\pm300$ & $7.0\pm0.5$ & $1$ &       & \phantom{$-$}$1$ & $1.3$ & $1310$ & \phantom{$-$}$-0.0130 \pm 0.0002$\phantom{$0$} & \phantom{$-$}$3460\pm 20$\phantom{$0$}\phantom{$0$} & \phantom{$-$}$0.290\pm 0.001$\phantom{$0$} & $\phantom{0}39$ & $\phantom{0}84$ & $\phantom{0}0.7$ & $\phantom{0}1.7$ & \phantom{$1$} & \phantom{$1$} \\
2710406 & $13.27$ & $6960\pm80$\phantom{$0$} & $12.6\pm0.9$ & $1$ &       & \phantom{$-$}$1$ & $1.5$ & $2270$ & \phantom{$-$}$-0.0054 \pm 0.0003$\phantom{$0$} & \phantom{$-$}$4270\pm 70$\phantom{$0$}\phantom{$0$} & \phantom{$-$}$0.138\pm 0.006$\phantom{$0$} & $\phantom{0}45$ & $\phantom{0}57$ & $\phantom{0}0.4$ & $\phantom{0}0.6$ & \phantom{$1$} & \phantom{$1$} \\
2710594 & $11.79$ & $7200\pm200$ & $8.0\pm0.5$ & $1$ &       & \phantom{$-$}$1$ & $0.7$ & $360$ & \phantom{$-$}$-0.0290 \pm 0.0003$\phantom{$0$} & \phantom{$-$}$4000\pm 10$\phantom{$0$}\phantom{$0$} & \phantom{$-$}$0.9920\pm 0.0006$ & $\phantom{0}27$ & $\phantom{0}86$ & $\phantom{0}2.3$ & $\phantom{0}7.9$ & \phantom{$1$} & \phantom{$1$} \\
        &         &         &         &       & $-2$ & $-1$ & $1.2$ & $470$ & $0.0381 \pm 0.0005$\phantom{$0$} &         &         & $\phantom{0}20$ & $\phantom{0}64$ & $\phantom{0}7.4$ & $18.0$ & \phantom{$1$} & \phantom{$1$} \\
2719928 & $12.66$ & $7200\pm200$ & $9.6\pm0.7$ & $2$ &       & \phantom{$-$}$2$ & $0.3$ & $70$ & \phantom{$-$}$-0.0275 \pm 0.0004$\phantom{$0$} & \phantom{$-$}$3700\pm 700$\phantom{$0$} & \phantom{$-$}$1.47\pm 0.04$\phantom{$0$}\phantom{$0$} & $\phantom{0}55$ & $\phantom{0}82$ & $\phantom{0}3.3$ & $\phantom{0}5.0$ & \phantom{$1$} & \phantom{$1$} \\
        &         &         &         & $1$ &       & \phantom{$-$}$1$ & $0.5$ & $150$ & \phantom{$-$}$-0.006 \pm 0.004$\phantom{$0$}\phantom{$0$} &         &         & $\phantom{0}59$ & $\phantom{0}69$ & $\phantom{0}7.1$ & $\phantom{0}8.4$ & \phantom{$1$} & \phantom{$1$} \\
2846358 & $11.02$ & $6800\pm200$ &        & $2$ &       & \phantom{$-$}$2$ & $0.4$ & $360$ & \phantom{$-$}$-0.0279 \pm 0.0004$\phantom{$0$} & \phantom{$-$}$4020\pm 50$\phantom{$0$}\phantom{$0$} & \phantom{$-$}$0.755\pm 0.004$\phantom{$0$} & $\phantom{0}25$ & $\phantom{0}61$ & $\phantom{0}0.7$ & $\phantom{0}2.0$ & $1$ & \phantom{$1$} \\
        &         &         &         & $1$ &       & \phantom{$-$}$1$ & $0.8$ & $650$ & \phantom{$-$}$-0.0336 \pm 0.0004$\phantom{$0$} &         &         & $\phantom{0}32$ & $\phantom{0}63$ & $\phantom{0}1.9$ & $\phantom{0}4.0$ & \phantom{$1$} & \phantom{$1$} \\
\hline
 \end{tabular} 
 \end{table} 
\end{landscape}

\subsection{Mode identification}\label{subsec:mode_identi}
\begin{figure}
    \centering
    \includegraphics[width=1\linewidth]{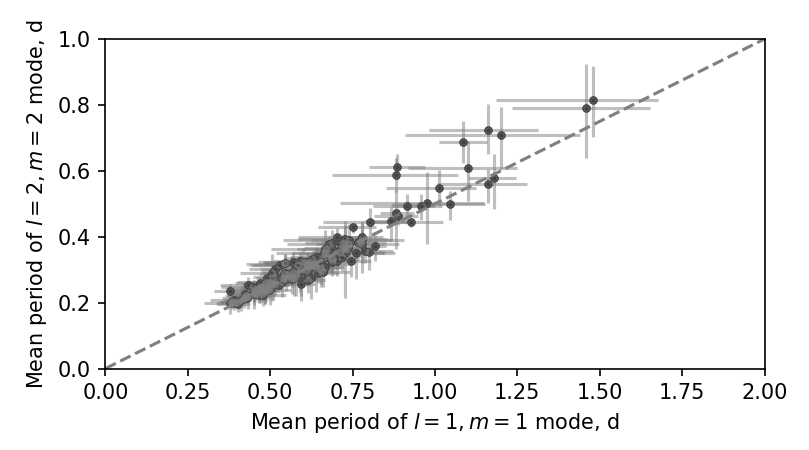}
    \caption{The observed period relation between $l=2,~m=2$ g modes and $l=1,~m=1$ g modes. The error bars show the pulsation period spans, not the uncertainties. The dashed line shows the relation that the mean period of $l=2,~m=2$ g modes is half that of $l=1,~m=1$ g modes.}
    \label{fig:l2_l1_period}
\end{figure}

\begin{figure}
    \centering
    \includegraphics[width=1\linewidth]{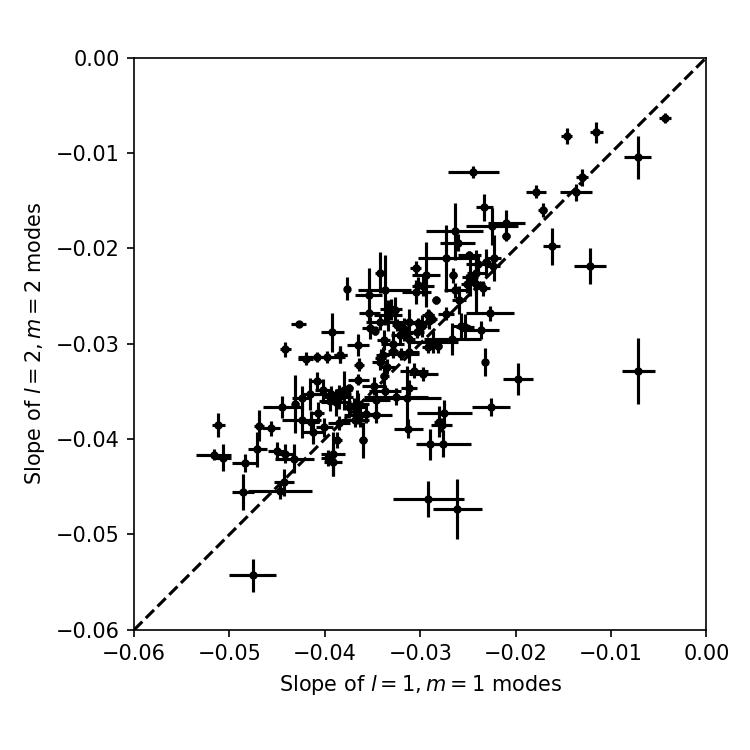}
    \caption{The slope relation between the observed $l=2,~m=2$ and $l=1,~m=1$ modes. We only plot the points with slope error smaller than 0.005. The dashed line shows the place where the slopes are equal. }
    \label{fig:l2_l1_slope}
\end{figure}

\begin{figure}
    \centering
    \includegraphics[width=1\linewidth]{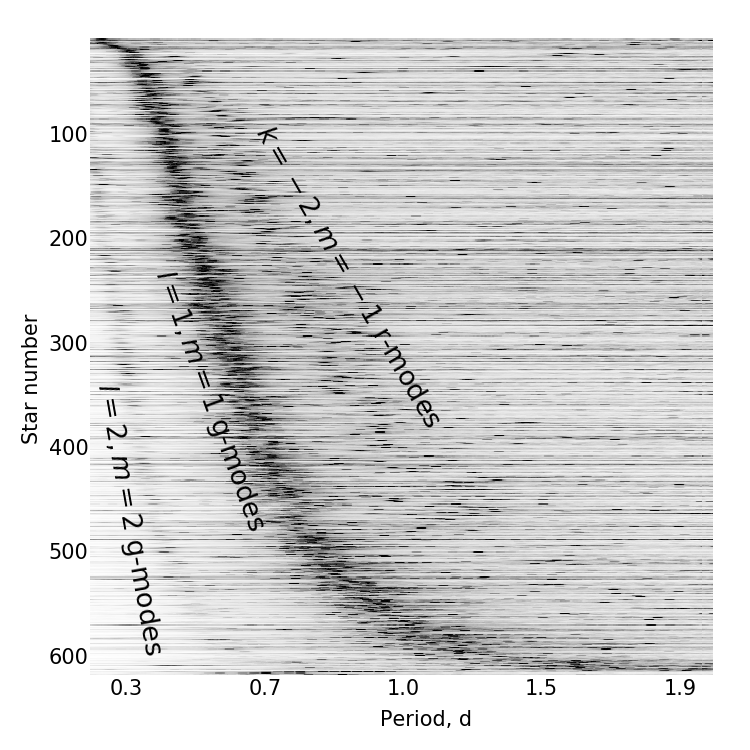}
    \caption{The periodograms of all the \gdor stars with identified period-spacing patterns in our sample. Each row shows the periodogram for one star. The colour stands for the normalised amplitudes to the power of 0.25 for the best visibility. The trends of $l=1,~m=1$ g modes, $l=2,~m=2$ g modes, and $k=-2, m=-1$ r modes are seen and marked by the texts. }
    \label{fig:periodogram_heatmap}
\end{figure}

The periodogram of a \gdor star generally shows peak groups which overlap with the harmonics of fundamental frequencies. We accepted the explanation by \cite{Saio_2018} that the peak groups are prograde sectoral g-mode oscillations of increasing angular degree. Many quadrupole modes are seen in our sample. Figure~\ref{fig:l2_l1_period} shows the correlation between the mean periods of $l=2,~m=2$ and $l=1,~m=1$ g modes. We find that the mean periods of quadrupole sectoral g modes are typically half those of dipole sectoral g modes, since the quadrupole sectoral g modes generally coincide with the second harmonics of dipole modes.

Figure~\ref{fig:l2_l1_slope} shows the slope relation between dipole and quadrupole g modes. We find that the slopes of $l=2,~m=2$ g modes are similar but slightly smaller than those of $l=1,~m=1$ g modes. We therefore conclude two features of the quadrupole sectoral g modes in \gdor stars:
\begin{itemize}
    \item the mean period of the quadrupole modes is half that of the dipole sectoral g modes.
    
    \item the slopes of quadrupole and dipole sectoral ($l=m$) g modes are almost equal.
\end{itemize}
These features are common in most of stars and help identify the modes. If not, several conditions should be considered: if the power spectrum is contaminated by a binary, or if they are $m=1$ and $m=0$ modes showing large splittings (see the example and discussion in Section~\ref{sec:fast_splitting}).

We plot the periodograms of all the \gdor stars in Fig.~\ref{fig:periodogram_heatmap} Each row displays the normalised periodogram of one star, sorted vertically by the mean period of the dipole modes. Three ridges are seen: the dominant one is $l=1,~m=1$ g modes; the ridge of $l=2,~m=2$ g modes appears on the left, as these modes overlap with the second harmonics of the dipole modes; the third ridge is the $k=-2, m=-1$ r modes.
We see that the $l=1,~m=1$ g modes in \gdor stars generally show the largest amplitudes. Assuming the dipole sectoral g modes appear around period of $P$, the quadrupole g modes are expected to appear around $0.5P$ and the r modes are more likely to appear around $2P$ \citep{Li_2019}. This structure of \gdor periodograms helps guide the mode identification. 

\textcolor{black}{We find that four stars (KIC5876187, KIC9344493, KIC10091792, KIC10803371) show $l\geq3$ g modes. These high $l$ g modes have smaller amplitudes and generally have periods below the lower boundary of our detection region (0.2\,d to 2\,d) hence they are hard to detect. However, for dipole and quadrupole g modes, we confirm that the results listed in 
Table~\ref{tab:rot_Pi0_table} are complete. }

\subsection{Occurence rate of modes}\label{subsec:percentage}

\begin{figure}
    \centering
    \includegraphics[width=1\linewidth]{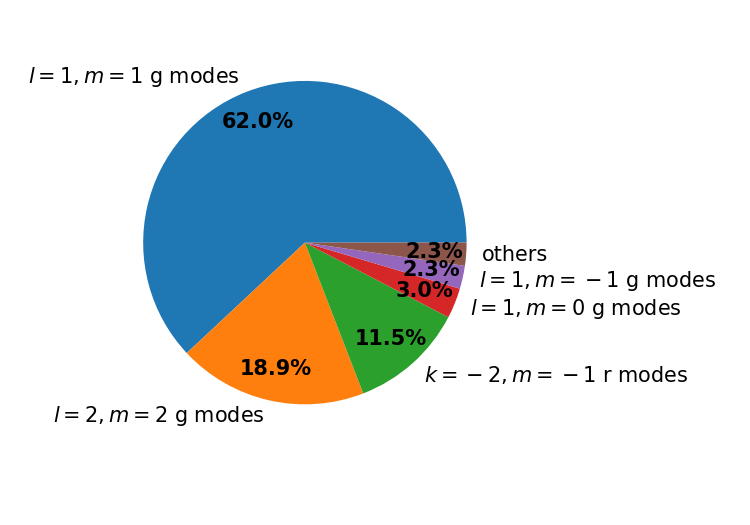}
    \caption{The percentages of different oscillation modes among \patternnumber patterns. }
    \label{fig:percentage}
\end{figure}

\begin{figure}
    \centering
    \includegraphics[width=1\linewidth]{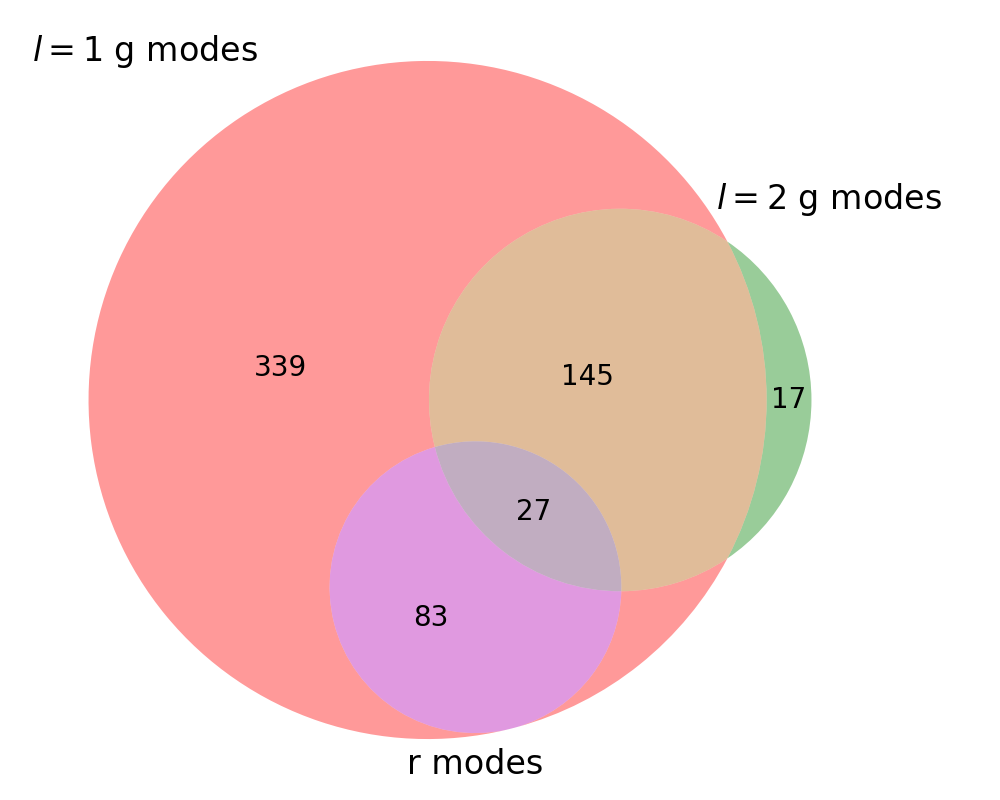}
    \caption{The numbers of stars with observed period-spacing patterns.}
    \label{fig:percentage_star}
\end{figure}

Figure~\ref{fig:percentage} shows the observed relative occurrence rates of different modes. Among all the \patternnumber patterns, 62.0\% are $l=1,~m=1$ g modes. The second most common modes are $l=2,~m=2$ g modes, which constitute 18.9\% of the total detection. Rossby modes are the third most common modes (11.5\%). Apart from these three modes, we also see $l=1,~m=0$ and $l=1,~m=-1$ modes with percentages of 3.0\% and 2.3\%, respectively, which were mainly found in the slow rotators reported by \cite{Li_2018}. \numberfastsplitting fast rotators with splittings are detected in this work, which will be described in Section~\ref{sec:fast_splitting}.

There are a few patterns that cannot be classified into those five types of modes above. They might be the sectoral g modes with higher angular degree ($l=3, m=3$ for example, see KIC\,9344493), or the only $k=-1, m=-1$ r mode reported by \cite{Li_2019}, or $l=2,~m=1$ g modes in two newly-discovered fast rotators (KIC\,5092681 and KIC\,5544996), or some patterns that cannot be fitted with the TAR. All of these occupy 2.3\% of the total detected period-spacing patterns.

Figure~\ref{fig:percentage_star} shows the numbers of stars which show different oscillation modes. We classify the modes into three main types: $l=1$ g modes, $l=2$ g modes, and $k=-2, m=-1$ r modes. There are 339 stars that only show $l=1$ g modes (red area) and 145 stars that show both $l=1$ and $l=2$ g modes (yellow area). Almost all the stars have $l=1$ g-mode period-spacing patterns. However, there are power excesses over $l=1$ g-modes regions in 16 stars without period-spacing patterns identified. For these stars only $l=2$ g-modes patterns are reported. We notice that KIC\,5491390 is the only star that does not show any $l=1$ g-mode power excess. In total, there are 17 stars with only $l=2$ g modes (green area). \cite{Zhang_2018} reported that KIC\,10486425 also oscillates only in $l=2$ g modes. The reason for the absence of $l=1$ g modes needs further investigation.

We do not find any star that only shows r modes. The reason is that the TAR cannot converge well if g-mode patterns are absent, hence we cannot ensure that the observed pattern is a real r-mode pattern, or we are misled by missing peaks in the observed pulsation spectra. Hence, all the r modes co-exist with g modes in our sample. There are 83 stars with $l=1$ g modes and r modes, and there are 27 stars with $l=1$, $l=2$ g modes, and r modes. The co-existence of g mode and r modes decreases the uncertainties of near-core rotation rates significantly. The typical uncertainty is 0.0009\,\cpd for the stars with both g and r modes, while it is 0.008\,\cpd for the stars with only g modes.

\subsection{Slope--Period diagram}
\begin{figure*}
\centering
\includegraphics[width=1\linewidth]{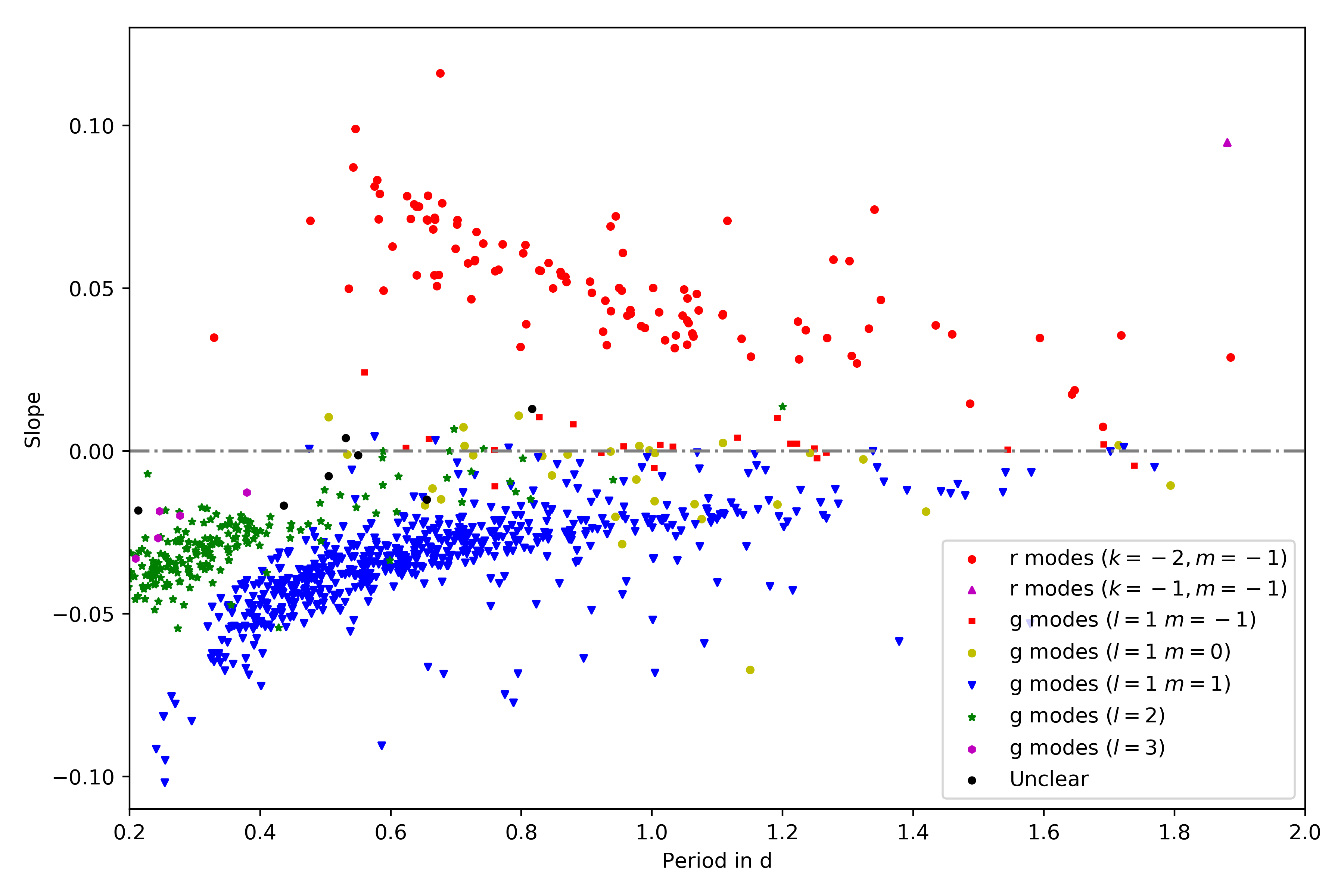}
\caption{Slopes vs the means of the periods of \patternnumber period-spacing patterns from \starnumber \gdor stars. The x-axis is the mean period in each pattern. The y-axis is the slope between the period spacing and the period with dips removed. Different colours and symbols show different modes. }\label{fig:period_slope}
\end{figure*}

In Section~\ref{sec:data_reduction}, we introduced three observables for each pattern, the mean period, the mean period spacing, and the slope. Figure~\ref{fig:period_slope} shows the relation between the slopes and the mean periods from all the patterns in our sample, hence we call this diagram the Slope--Period (S--P) diagram. The mean period and the slope are correlated. We find that the data points of $l=1,~m=1$ g modes, $l=2,~m=2$ g modes, slowly-rotating g modes, and $k=-2, m=-1$ r modes form four different groups which have diverse trends and clear boundaries.
\begin{itemize}
	\item $l=1,~m=1$ g modes: these points are the majority, which are shown by the blue triangles. Most patterns have mean periods between 0.4\,d and 0.8\,d and slopes around $\sim-0.04$. They show a positive correlation between the slopes and the mean periods.
	
	\item $l=2,~m=2$ g modes: those points are presented by the green stars. These modes have shorter periods than dipole modes (between 0.2\,d and 0.4\,d) but have similar slopes ($\sim -0.04$) as pointed out in Section~\ref{subsec:mode_identi}. 
	
	\item  $l=1,~m=0$ and $l=1,~m=-1$ g modes: they are marked by the yellow circles and the red rectangles. These two modes are rare (for $m=0$) or absent (for $m=-1$) in rapid rotators but are seen in the slow rotators reported by \cite{Li_2018}. Due to the slow rotation rate, the rotational effect is not strong so the period spacings in those modes remain nearly identical. Hence we see most $m=0$ and $m=-1$ modes around the horizontal line with slope of zero.
	
	\item $k=-2, m=-1$ r modes: they are the red circles. As discussed by \cite{Li_2019}, r modes have positive slopes and show an inverse correlation between the mean period and the slope. 
\end{itemize}

Figure~\ref{fig:period_slope} displays the typical locations of different modes on the S--P diagram. It can be used for mode identification. When a new pattern is found, its location on the S--P diagram reveals its mode identification. If the point is an outlier, several possibilities should be considered: the period spacings are misidentified since some peaks in the amplitude spectra are too weak to be detected; the slope is strongly affected by the partially-observed dips caused by chemical composition gradients \cite[e.g. KIC\,4919344 in][]{Li_2018}; or the star is a Slowly Pulsating B (SPB) star, which has a larger asymptotic spacing ($\Pi_0$) because it has a higher mass than a \gdor star \citep[e.g.][]{Papics_2017}. Consequently, a pattern of an SPB star has a steeper slope than a pattern of a \gdor star with a similar mean period.

\section{Asymptotic spacing and rotation}\label{sec:TAR_results}

We used the TAR to fit the period-spacing patterns and measured the near-core rotation rates \frot, the asymptotic spacings $\Pi_0$ (also called buoyancy radii), and the radial orders $n$. 

\subsection{Distribution of $\Pi_0$}

\begin{figure}
\centering
\includegraphics[width=1\linewidth]{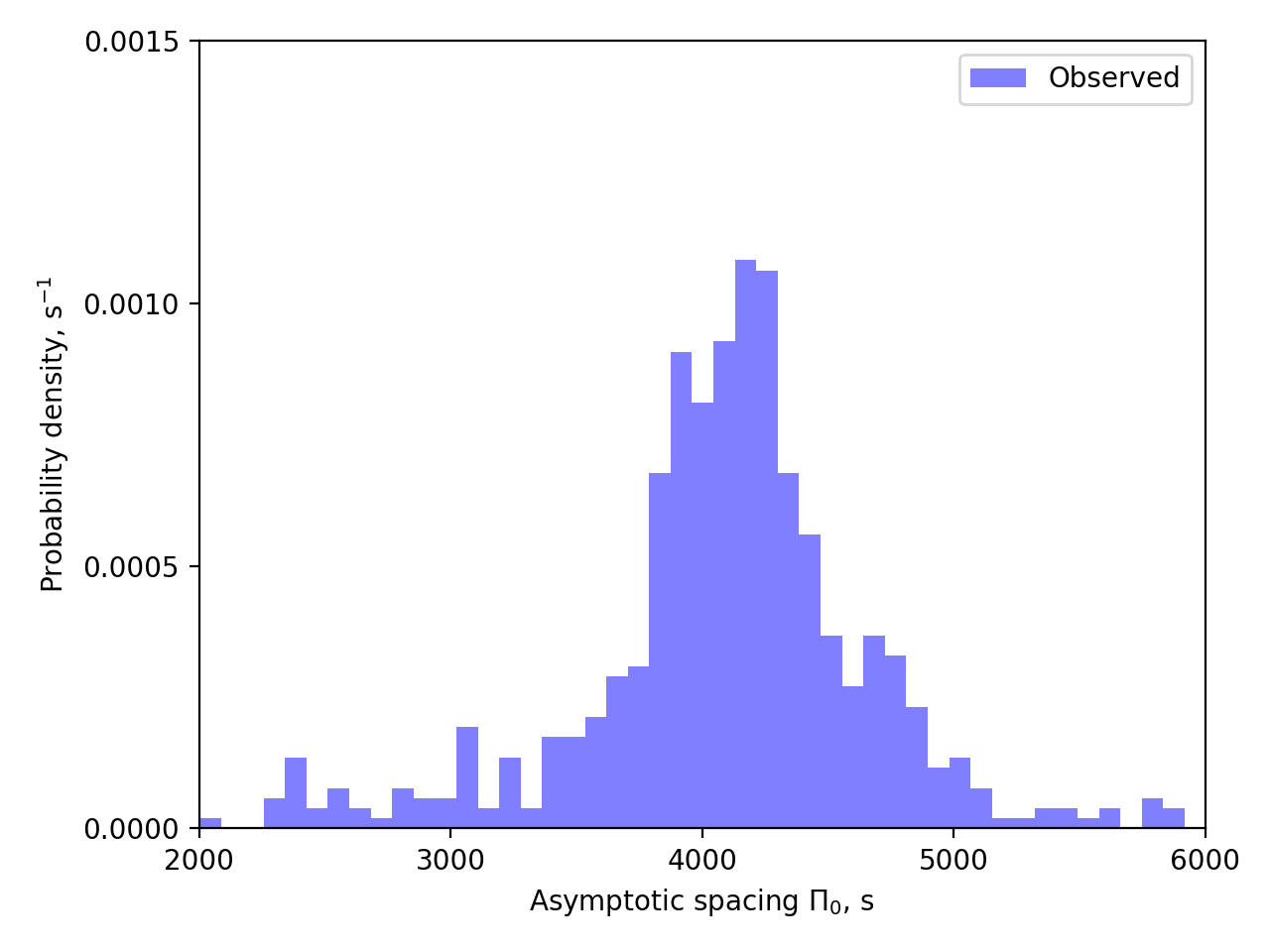}
\caption{The observed distribution of asymptotic spacing $\Pi_0$. }\label{fig:Pi0_hist}
\end{figure}

\begin{figure}
    \centering
    \includegraphics[width=\linewidth]{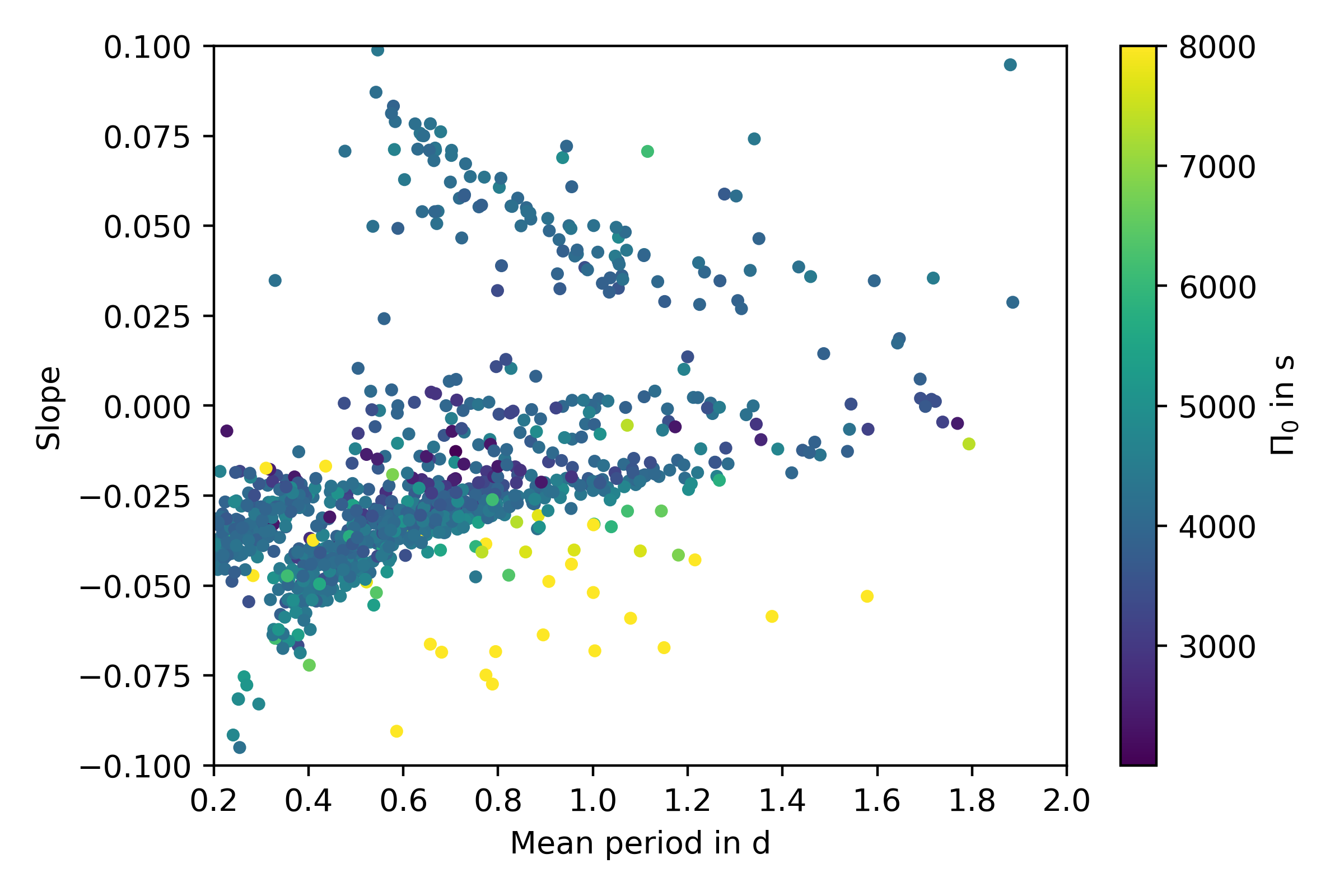}
    \caption{The S--P diagram coloured by the asymptotic spacing $\Pi_0$. The outliers on the lower right is composed of the stars with large $\Pi_0$. }
    \label{fig:SP_diagram_Pi0}
\end{figure}

\begin{figure}
\centering
\includegraphics[width=1\linewidth]{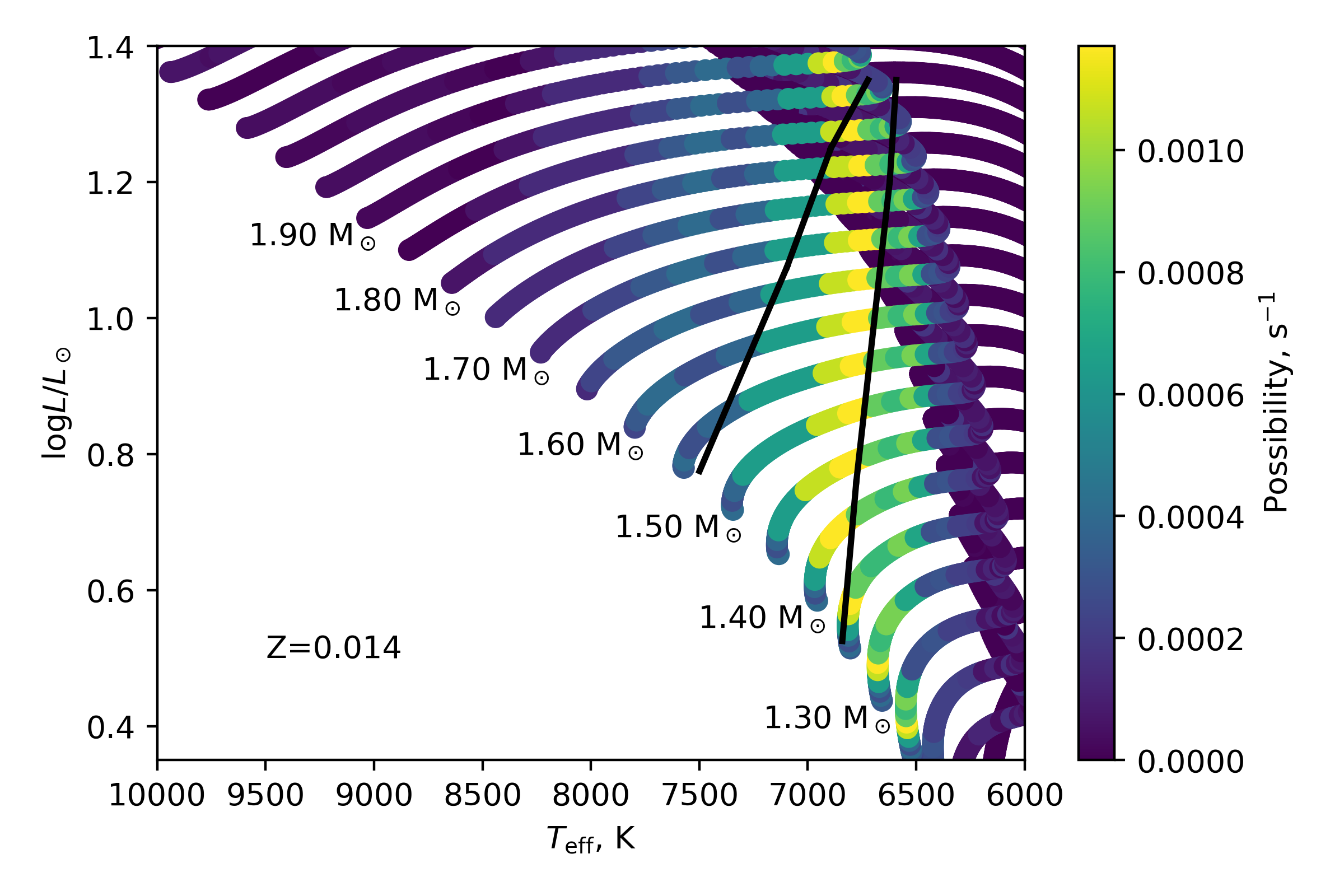}
\caption{The probability density of $\Pi_0$ on the HR diagram. The colour stands for the probability density of $\Pi_0$ from Fig.~\ref{fig:Pi0_hist}. The black lines show the theoretical instability strip of \gdor stars \citep{Dupret_2005}. }\label{fig:Pi0_HR_diagram}
\end{figure}

\begin{figure}
\centering
\includegraphics[width=1\linewidth]{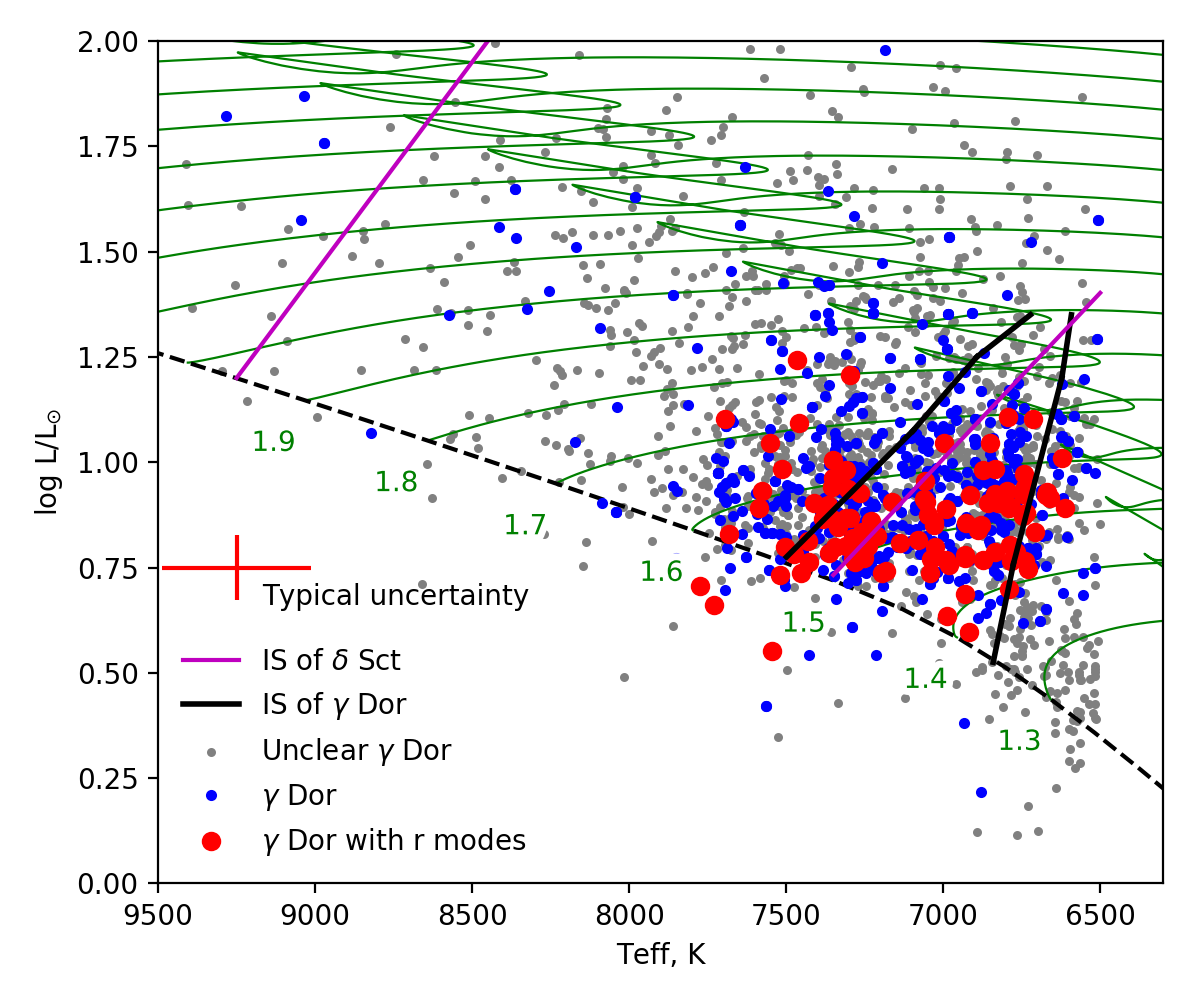}
\caption{Stars on the HR diagram. The blue points are the \gdor stars with clear g-mode patterns. The red points are the stars with both g- and r-mode patterns. The grey points are the stars we inspected but do not show clear period spacing patterns. The black lines show the theoretical IS of \gdor star while the purple lines depict the observed IS of $\delta$ Sct star by \protect\cite{Murphy_2019}. }\label{fig:stars_on_HR_diagram}
\end{figure}

Figure~\ref{fig:Pi0_hist} gives the observed distributions of the asymptotic spacing $\Pi_0$. The stars show a symmetric distribution centred around $\Pi_0=4000\,\mathrm{s}$. We find that 68\% of stars have $\Pi_0$ between 3700\,s and 4800\,s. Stars with large $\Pi_0$ are likely to be Slowly Pulsating B (SPB) stars \cite[with $\Pi_0$ from 5600\,s to 16000\,s][]{Papics_2017}. They show g-mode patterns with larger period-spacing values and steeper slopes than \gdor stars but the pulsation periods are similar. However, the effective temperatures of those possible SPB stars are located in the typical ranges of A- and F-type stars. This may indicate that the effective temperatures are wrong, or there are pulsation periods missing in the detected patterns, or the stars are very young. The stars with $\Pi_0\lesssim 3000\,\mathrm{s}$ are probably close to the terminal age main sequence. 

\cite{VanReeth_2016_TAR} reported the theoretical distribution of $\Pi_0$, which was calculated based on a grid of theoretical stellar models that includes the \gdor instability strip. The relative duration of the different evolutionary stages was also considered in the calculation. It shows that the most likely value of $\Pi_0$ is 4400\,s, which is higher than the observed one. The theoretical histogram also has a slightly asymmetric shape. The discrepancy between the observed and theoretical distributions are probably caused by the different parameters, such as metallicity and mixing length, and it also reveals that a full non-adiabatic computation of the \gdor instability strip is needed for the theoretical distribution of $\Pi_0$.

Figure~\ref{fig:SP_diagram_Pi0} shows the S--P diagram coloured by their asymptotic spacings $\Pi_0$. We found that the yellow outliers on the lower right are composed of the stars with large $\Pi_0$. They generally show steeper period-spacing patterns hence they appear below the typical $l=1,~m=1$ g-mode group of \gdor stars. 
From now, we only present the results using stars with $\Pi_0<6000\,\mathrm{s}$ to avoid any contamination from SPB stars or wrong identifications.

We show our theoretical evolutionary tracks in Fig.~\ref{fig:Pi0_HR_diagram}. \textsc{MESA} v10108 was used to compute the evolutionary tracks \citep{Paxton_2011, Paxton_2013, Paxton_2015, Paxton_2018}. The tracks shown in Fig.~\ref{fig:Pi0_HR_diagram} have: stellar masses are from 1.0 to  $3.0\,\mathrm{M_\odot}$ with step of $0.05\,\mathrm{M_\odot}$, a hydrogen mass fraction $X$ of 0.71, a metallicity $Z$ of 0.014, a mixing length $\alpha$ of 1.8, an exponential core overshooting $f_\mathrm{ov}$ of 0.015, and an extra diffusive mixing $D_{\rm mix}$ of 1\,cm$^{2}$s$^{-1}$, we also used the OPAL capacities and the \citet{Asplund_2009} solar abundance mixture. For each stellar model, the asymptotic spacing $\Pi_0$ is calculated and the point is coloured by the observed probability density of $\Pi_0$ from Fig.~\ref{fig:Pi0_hist}. Two solid black lines in Fig.~\ref{fig:Pi0_HR_diagram} display the boundaries of the theoretical instability strip of \gdor stars \citep{Dupret_2005}. We find that the areas with high $\Pi_0$ densities show a nearly-vertical strip, broad at the ZAMS and narrow at the TAMS. The low-mass stars are more likely to pulsate near the ZAMS, while for the high-mass stars, the pulsation may happen close to the TAMS and for a shorter duration than for low-mass stars. The high-density area of $\Pi_0$ on the HR diagram is generally consistent with the theoretical instability strip. 

Combining the effective temperatures from \textit{Kepler} DR25 \citep{Mathur_2017}, and the luminosities from \cite{Murphy_2019} using \textit{gaia} DR2 parallax \citep{Prusti_2016}, we place our stars on the HR diagram, as shown in Fig.~\ref{fig:stars_on_HR_diagram}. Figure~\ref{fig:stars_on_HR_diagram} displays that most \gdor stars are located on the lower right area, with lower effective temperature and luminosity than $\delta$\,Sct stars. The low-temperature boundary of our \gdor sample follows the theoretical instability strip (solid black lines). This may prove that the theory predicted the red boundary correctly. However, many \gdor stars are located beyond the blue boundary of the instability strip. 
This could be caused by systematic offsets in the photometric $T_\mathrm{eff}$ values. Typical uncertainties on these values are on the order of 250\,K. More accurate $T_\mathrm{eff}$ values from high-resolution spectroscopy are needed to evaluate this possibility. If the $T_\mathrm{eff}$ values are found to be accurate, the presence of hot \gdor stars could reflect the limit of the current theory, which was mentioned by \cite{Dupret_2005}. For example, a proper mixing length should be used in these stars rather than the solar value. 

As mentioned before, we inspected \totalnumber stars and found \starnumber stars with clear period-spacing patterns. The grey circles in Fig.~\ref{fig:stars_on_HR_diagram} are the stars without identified period-spacing patterns. They may be the \gdor stars with unresolved g-mode patterns, or the phase-modulation binaries from \cite{Murphy_2018}. We included the phase-modulation binaries since they show similar $T_\mathrm{eff}$, but they may not necessarily be \gdor stars. We do not find any special distributions of the stars without pulsation patterns on the HR diagram. There is no explanation about why some \gdor stars do not show any clear period-spacing pattern. The reasons might be: dense, overlapping patterns that would be hard to disentangle; there are only a few excited modes hence the pattern is incomplete.

\subsection{Distribution of \frot with slow-rotator excess}\label{subsec:bimodality_frot}

\begin{figure}
\centering
\includegraphics[width=1\linewidth]{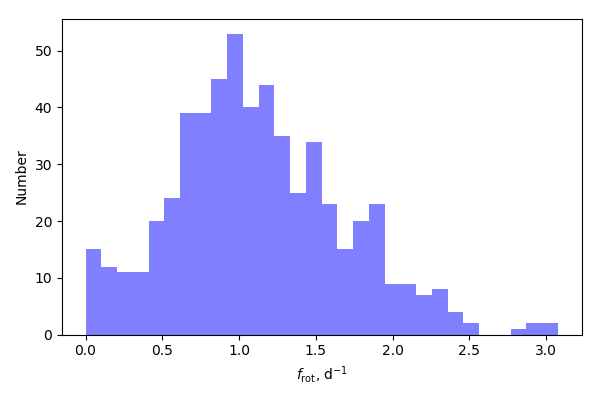}
\caption{The distribution of observed near-core rotation rate $f_\mathrm{rot}$. Many stars rotate around $1$\,$\mathrm{d^{-1}}$ while there is a slow-rotator excess slower than 0.4\,$\mathrm{d^{-1}}$, suggesting two classes of \gdor stars. }\label{fig:frot_hist}
\end{figure}

\begin{figure*}
\centering
\includegraphics[width=1\linewidth]{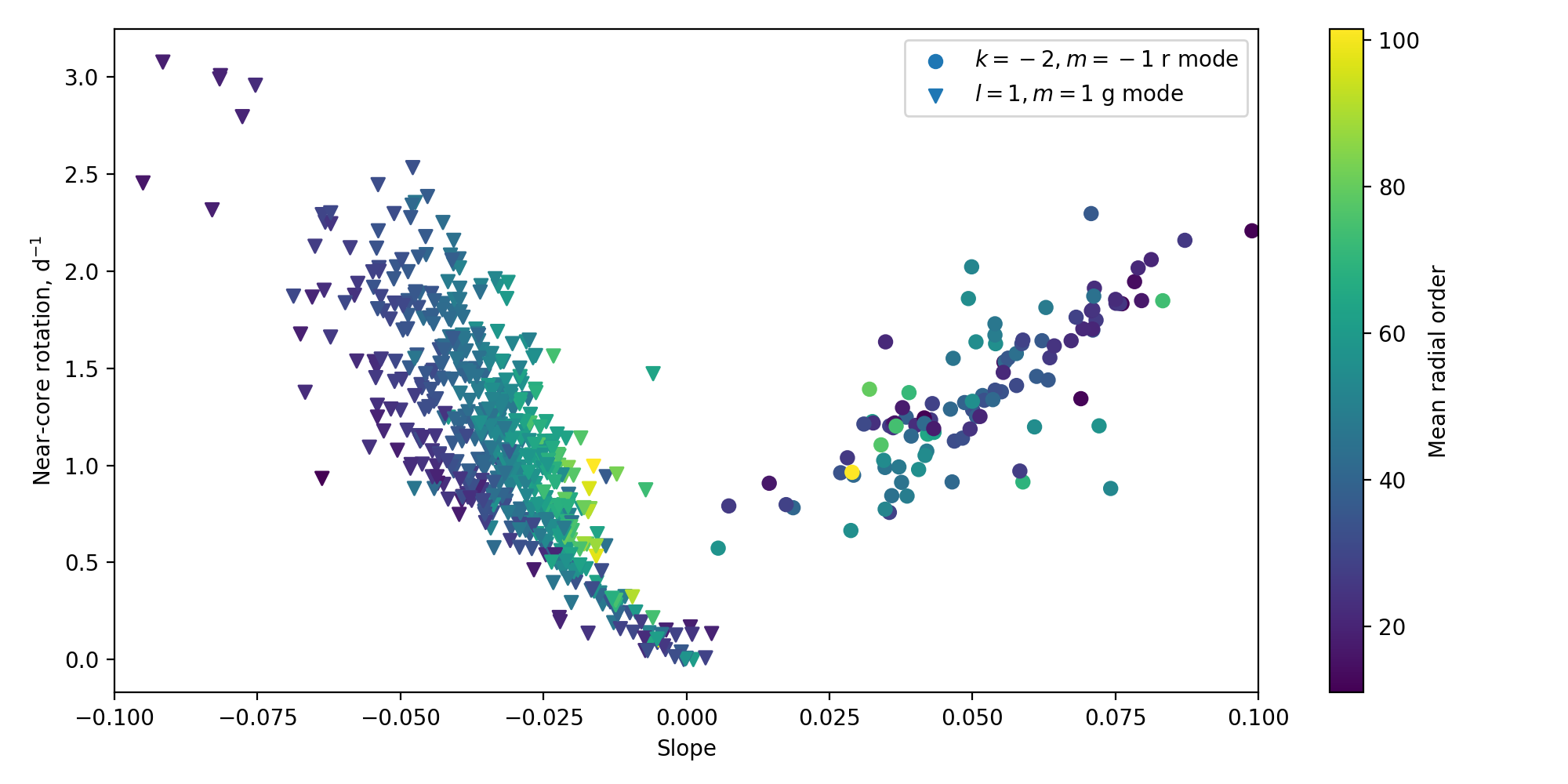}
\caption{The relation between slope and near-core rotation rate $f_\mathrm{rot}$. The triangles are the $l=1,~m=1$ g-mode patterns, whose slopes are generally smaller than zero hence are located on the left. The circles are the $k=-2, m=-1$ r-mode patterns with positive slopes on the right. The g-mode slopes are correlated with their mean radial orders, as shown by the colour gradient. }\label{fig:slope_rotation_radial_order}
\end{figure*}

\begin{figure}
    \centering
    \includegraphics[width=\linewidth]{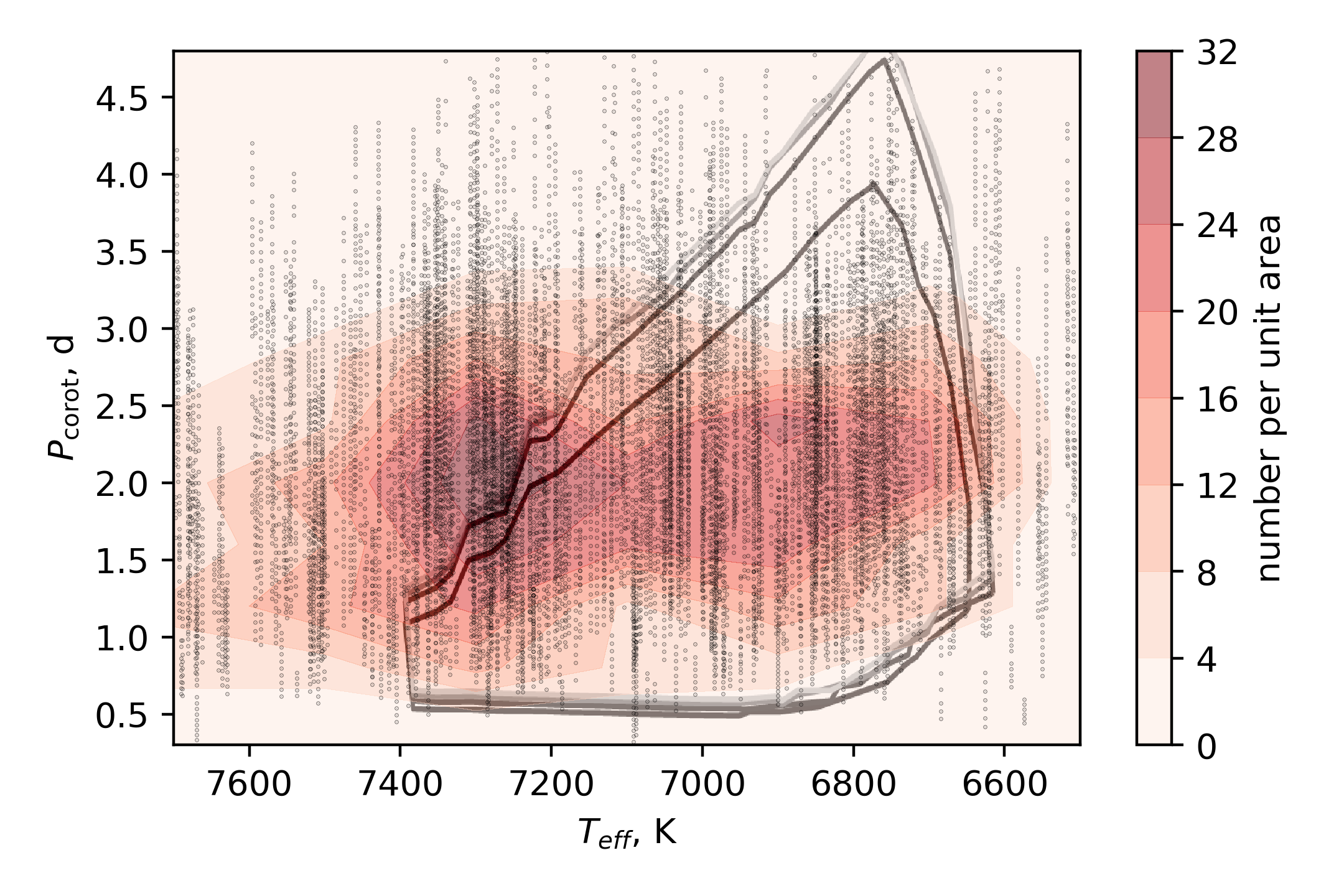}
    \caption{Pulsation period in the co-rotating frame vs effective temperature. One star has one effective temperature but many pulsation modes hence the data points (black dots) show vertical fringes. The contour displays the number of the observed pulsation modes. The solid lines show the theoretical areas reported by \protect\cite{Bouabid_2013}, with equatorial rotation velocities $v_\mathrm{eq}=0$ (black), 30 (dark grey), 60 (grey) and 90\,$\mathrm{km/s}$ (light grey). }
    \label{fig:Pcor_Teff}
\end{figure}

Figure~\ref{fig:frot_hist} displays the distribution of the near-core rotation rates \frot. Most stars have rotation frequencies around 1\,\cpd. The distribution increases rapidly after 0.4\,$\mathrm{d^{-1}}$ and drops slowly after $\sim1$\,$\mathrm{d^{-1}}$. The most rapid rotators are KIC\,8458690A and KIC\,8458690B with \frot$\sim3.01$\,$\mathrm{d^{-1}}$, whose two identical period-spacing patterns form `splittings' reported by \cite{Li_2018}. However, many stars rotate less quickly than expected, which forms an excess at \frot$\lesssim0.4$\,$\mathrm{d^{-1}}$  in Fig.~\ref{fig:frot_hist}.  

The histogram of the near-core rotation rate in Fig.~\ref{fig:frot_hist} shows a slow-rotator excess. We suggest defining two classes of \gdor stars by their near-core rotation rates: (1) slow rotators with $f_\mathrm{rot} \lesssim 0.4$\,$\mathrm{d^{-1}}$; (2) fast rotators with $f_\mathrm{rot} \gtrsim 0.4$\,$\mathrm{d^{-1}}$. A similar distribution has been realised for A- and F-type stars by observing the projected velocity $v \sin i$ \citep[e.g.][]{Ramella_1989, Abt_1995, Royer_2007}. \cite{Abt_1995} found that all the rapid rotators have normal spectra and nearly all slow rotators have abnormal spectra (Ap or Am). The extremely slow rotation rate may be explained by magnetic braking for Ap stars, or tidal braking for Am stars. However, after removing Ap and Am stars, \cite{Royer_2007} still found the bimodality. The slow rotators have $v \sin i < 70\,\mathrm{km/s}$ and the fast rotators have $v \sin i \sim 160\,\mathrm{km/s}$, whose ratio is consistent with our near-core rotation rate (0.4\,$\mathrm{d^{-1}}$ and 1.0\,$\mathrm{d^{-1}}$). Due to the large sample size here, the effect of inclination should be averaged out hence we compare $v \sin i$ with our inclination-independent near-core rotation rate in the last sentence directly. Rotational braking during the main sequence can be explained in many ways, such as magnetic fields, binarity, interaction with stellar disc, or the formation of blue stragglers \citep[e.g.][]{Mestel_1968, Hut_1981, Takada-Hidai_2017}. Our sample contains a large number of slow rotators. Follow-up spectroscopic observations can obtain the chemical abundances and the surface rotations, hence we can infer the formation of the slow rotators.

According to \cite{Ouazzani_2017}, the slope $\Sigma=\mathrm{d}\Delta P/ \mathrm{d}P$ was defined as a diagnostic for rotation. The slope decreases from zero with increasing rotation for $m\geq0$ g modes and vice versa for the r-mode pattern. We plot the relation between the fitted near-core rotation rate and the observed slope in Fig.~\ref{fig:slope_rotation_radial_order}. The points clustered into two groups, corresponding to $l=1,~m-1$ g modes and $k=-2, m=-1$ r modes. For the g-mode patterns, only several slow rotators show g-mode slopes slightly larger than 0 and most points have negative slopes and are located on the left side of Fig.~\ref{fig:slope_rotation_radial_order}. We find the rotation--slope relation of the g modes has a large scatter ($\sigma=0.35$\,$\mathrm{d^{-1}}$) and shows an obvious gradient with the mean radial orders. The gradient reveals that the slope for a period-spacing pattern is not only affected by the rotations and dips, but also affected by the radial orders. For a given rotation rate (for example 1\,$\mathrm{d^{-1}}$), the slopes are generally flatter ($\Sigma$ near zero) for higher radial orders. The effect of radial orders is clear and can be used to explain the widths of the trends in Fig.~\ref{fig:period_slope}. Further discussion about the radial orders on the S--P diagram will be given in Section~\ref{sec:rot_on_S_P_diagram} and Fig.~\ref{fig:iso_rotation}.

For the r modes, the rotation rate has a positive correlation with the slope. There is less scatter among the r-mode points ($\sigma$ = 0.24\,$\mathrm{d^{-1}}$), presumably because they show a smaller spread in radial orders, as we investigate in Section~\ref{subsec:radial_order_distribution}. The theoretical relation between slope and near-core rotation rate also depends on the stellar parameters (such as $T_\mathrm{eff}, \mathrm{[Fe/H]}$), which should be considered when comparing with observations.

\cite{Bouabid_2013} predicted the relation between the pulsation period in the co-rotating frame vs the effective temperature, shown as the solid lines in Fig.~\ref{fig:Pcor_Teff}. It predicted that \gdor stars pulsate between 0.5 and 5\,d in the co-rotating frame with effective temperature from 6600 to 7400\,K. The area is triangular, implying that the long-period stars are more likely to have lower temperatures. We count the number of the $l=1,~m=1$ g modes and compare our observations with the theoretical prediction in Fig.~\ref{fig:Pcor_Teff}. The pulsation period in the inertial frame is converted into the co-rotating frame using the near-core rotation rate derived by the TAR fit. It shows that the co-rotating periods are generally between 1 and 4\,d, following the theoretical prediction. However, many stars have higher photometric temperatures than the theory, which is similar to what the H-R diagram shows in Fig.~\ref{fig:stars_on_HR_diagram}. The peak of the observed contour is located outside the theoretical area and the pulsation period does not show any relation with effective temperature. 

\subsection{Correlation between $\Pi_0$ and \frot}

\begin{figure*}
\centering
\includegraphics[width=1\linewidth]{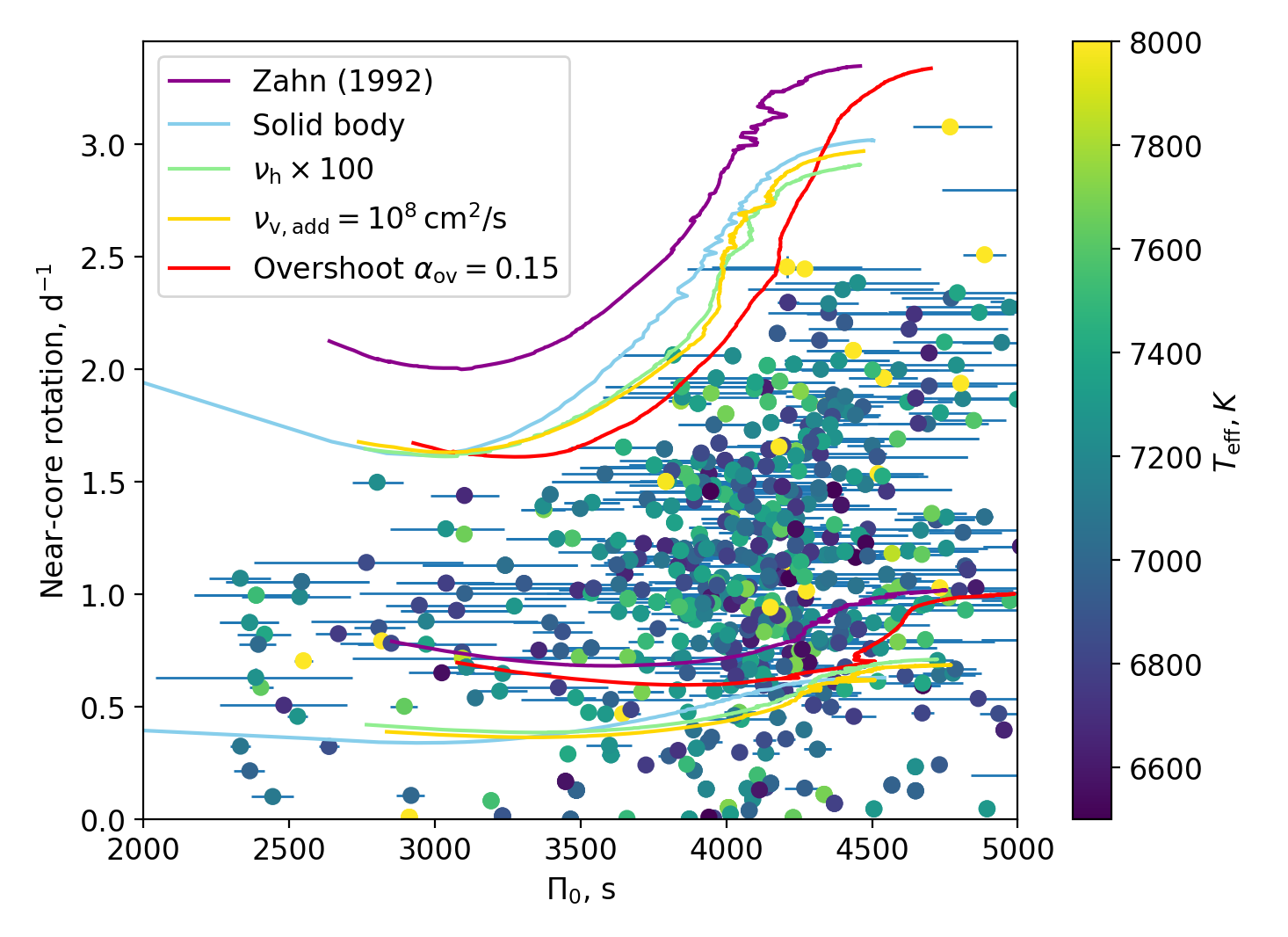}
\caption{The correlation between asymptotic spacing $\Pi_0$ and the near-core rotation rate $f_\mathrm{rot}$. With stellar evolution, the star moves from right to left. We only plot the points with $\Pi_0$ uncertainty smaller than 500\,s. The colour stands for the effective temperature. The theoretical predictions (solid lines) are made by \protect\cite{Zahn_1992} and \protect\cite{Ouazzani_2018}, where $\nu_\mathrm{h}$ means the enhanced horizontal viscosity, $\nu_\mathrm{v,add}$ means the additional vertical viscosity. }\label{fig:correlation}
\end{figure*}

Figure~\ref{fig:correlation} shows the correlation between the near-core rotation rate and the asymptotic spacing. The uncertainty on $\Pi_0$ is sometimes large, due to short patterns, or when only $l=2,~m=2$ g modes are seen. Hence, we only plot the 578 stars with $\Pi_0$ uncertainty within $\pm$500\,s. The asymptotic spacing decreases with stellar evolution so it is considered an indicator of the stellar age \citep[e.g.][]{Saio_2015, Ouazzani_2018}. However, the relation between $\Pi_0$ and age is affected by many other issues, for example, the shape of the instability strip (Fig.~\ref{fig:Pi0_HR_diagram}), or the initial mass. A detailed relation was reported by \cite{Mombarg_2019}. With stellar evolution, angular momentum is transferred and the near-core rotation also decreases. Hence, in Figure~\ref{fig:correlation}, the stars evolve from upper-right to lower-left.

The solid lines show the theoretical boundaries calculated by the angular momentum transfer model of \cite{Zahn_1992}, which considered the effects of meridional circulation and shear-induced turbulence and were calibrated by the observations of three clusters \citep{Ouazzani_2018}. We plot the theoretical boundaries with different conditions, such as the original model by \cite{Zahn_1992} (purple lines), the model assuming stars are solid bodies (blue lines), the \cite{Zahn_1992} model with enhanced horizontal viscosity $\nu_\mathrm{h}\times 100$ (green lines), or with additional vertical viscosity $\nu_\mathrm{v, add}$ (yellow lines), or with overshooting $\alpha_\mathrm{ov}=0.15$ (red lines) \citep[see details in][]{Ouazzani_2018}.

Our observational points are generally located between the theoretical lines. 
The upper boundaries of the models from \cite{Ouazzani_2018} fit the observations very well, which were calibrated by three clusters to include 80\% of the stars. Our observations do not have any star above these upper boundaries, implying that there is a lack of fast rotators.
We also notice that there are still many slow rotators below the lower boundaries of these models, confirming the `slow rotator accumulation' by \cite{Ouazzani_2018}. 
Our results are consistent with and expand upon the results from 37 stars by \cite{Ouazzani_2018} (these stars are also in our sample). 

The difference between the observations and the theory demands an explanation. Either there is a selection effect in the observations, or there are ingredients missing from the stellar models that produce the theoretical predictions. We consider the former, first.

The `fast rotators desert' might be expected if the period spacing patterns of rapid rotators cannot be extracted from (evolved) stars with small asymptotic spacings, as is indeed the case. In other words, although patterns are extractable for $\Pi_0 = 5000$\,s and $f_{\rm rot} = 2.5$\,d$^{-1}$, they are not extractable for $\Pi_0 = 3000$\,s at the same $f_{\rm rot}$ due to a denser power spectrum. However, the shape of the theoretical regions in Fig.~\ref{fig:correlation} matches the observational distribution and is only shifted from it. Stars are not predicted at $\Pi_0 = 3000$\,s and $f_{\rm rot} = 2.5$\,d$^{-1}$, so the `fast rotators desert' is not the result of an observational selection effect.

What ingredients might be missing from the models that would move the theoretical region towards the observed one? One possible answer is the rigid rotation. As pointed out by \cite{VanReeth_2018, Li_2019} and discussed in Section~\ref{sec:surface_rotation}, the \gdor stars have almost the same rotation rates between near-core and surface regions, implying a very effective mechanism of angular momentum transfer. The model with solid body condition (light blue lines in Fig.~\ref{fig:correlation}) indeed shifts down and is a better match to the observations.
\citet{Ouazzani_2018} also modified different coefficients beyond their ordinary range to investigate the effect of the models, such as enhancing horizontal viscosity in the star by a factor 100, which also have the desired effects. 
Another governing variable is the asymptotic spacings \citep[or called `buoyancy radius' in][]{Ouazzani_2018}; increasing the asymptotic spacings moves the theoretical region down in Fig.~\ref{fig:correlation}. An additional parameter that modifies the asymptotic spacings is convective overshooting above the core. Including this parameter is physically motivated, it migrates the theoretical boundaries in the direction of the observations, and may fully resolve the difference between the theory and observations, as what we see in Fig.~\ref{fig:correlation}.


\subsection{Distributions of radial orders}\label{subsec:radial_order_distribution}
\begin{figure}
    \centering
    \includegraphics[width=\linewidth]{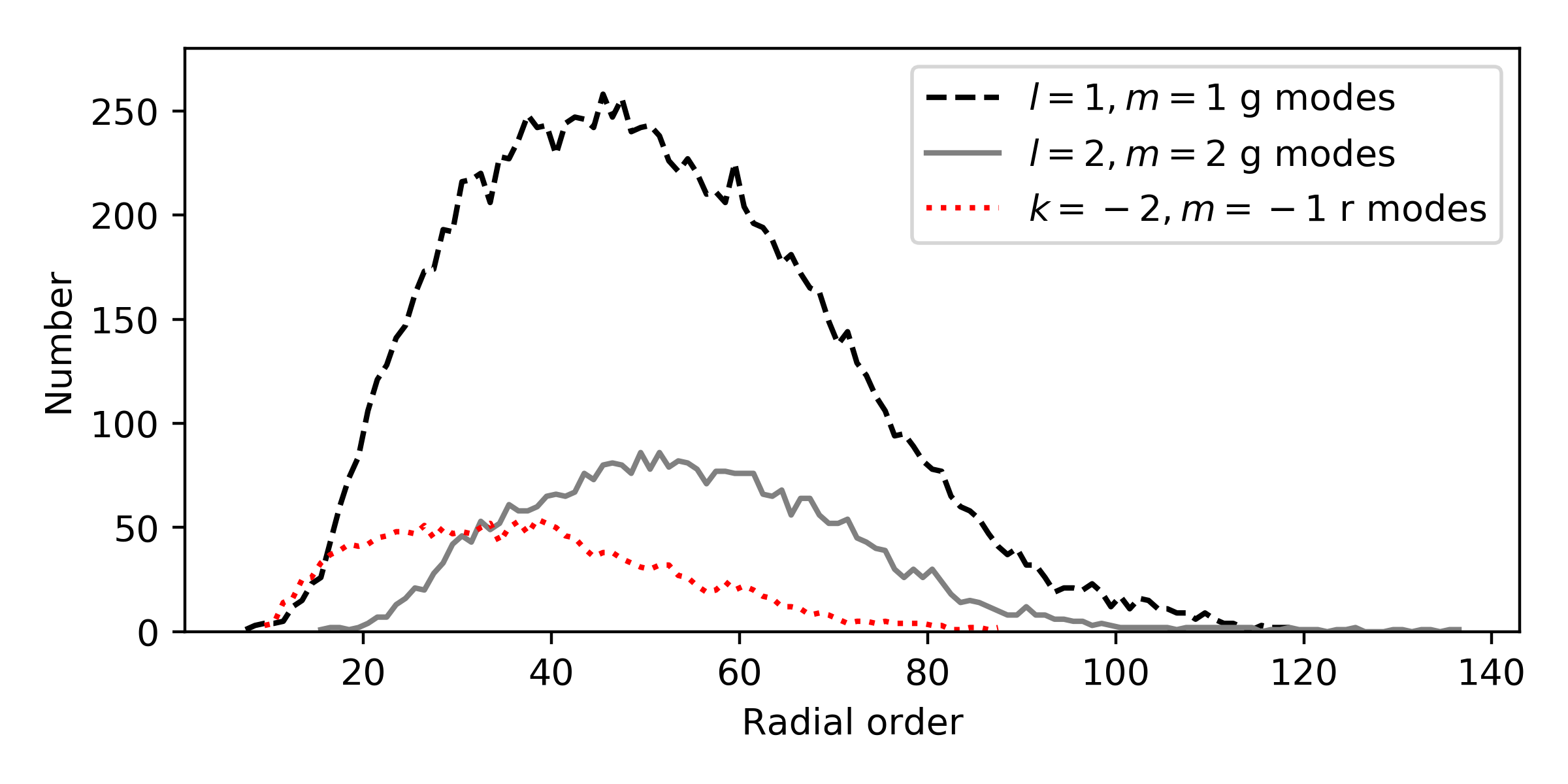}
    \caption{The distributions of radial orders of different modes.}
    \label{fig:radial_order_distribution}
\end{figure}

\begin{figure}
    \centering
    \includegraphics[width=\linewidth]{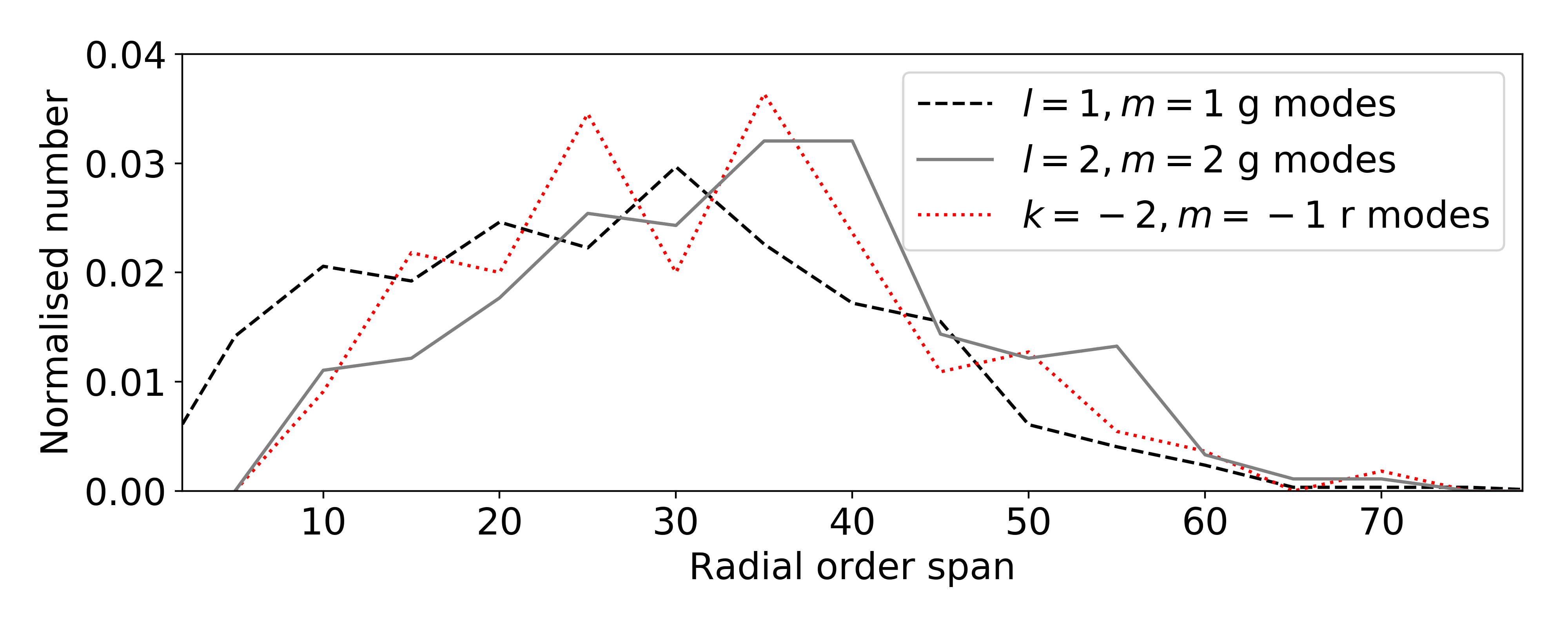}
    \caption{The normalised distributions of the pattern lengths.}
    \label{fig:length_distribution}
\end{figure}

 Figure~\ref{fig:radial_order_distribution} depicts the distributions of the radial orders for different modes, which are obtained by the best-fitting results of the TAR. We find that the distributions of the radial orders are similar to the results by \cite{Li_2019}. For $l=1,~m=1$ g modes, the median of the distribution is 48, and 68\% of modes have radial orders between 30 and 70. For $l=2,~m=2$ g modes, the peak of the distribution has a slightly higher radial order than dipole g modes, and 68\% of modes satisfy $37<n<71$. We notice that the radial orders are higher than the theoretical prediction by \cite{Bouabid_2013}, which found that the modes with radial orders from 15 to 38 are unstable.
 
For $k=-2, m=-1$ r modes, the radial orders are generally lower than those of g modes. The median is 36, and $21<n<53$ is the range for 68\% of the modes. The distribution of r-mode radial orders is asymmetric while the distributions of g modes are almost symmetric.

Figure~\ref{fig:length_distribution} presents the pattern lengths for different modes. The pattern length is the difference between the maximum and minimum of the radial orders. The numbers are normalised for a clear comparison. We find that the lengths for dipole g modes, quadrupole g modes, and r modes do not show any dramatic differences. The medians are about 30 and most of them have pattern lengths between 10 and 50 radial orders. Several patterns are extremely long, even up to 70 radial orders.

\cite{Bouabid_2013} calculated the radial order span using the theory of mode stability and found the radial order span is typically 30. Our observed radial order span is longer than the theory, showing that improvement of the mode excitation and damping theory may be needed.

\subsection{Distributions of spin parameters}\label{subsec:spin_para_dist}

\begin{figure}
    \centering
    \includegraphics[width=\linewidth]{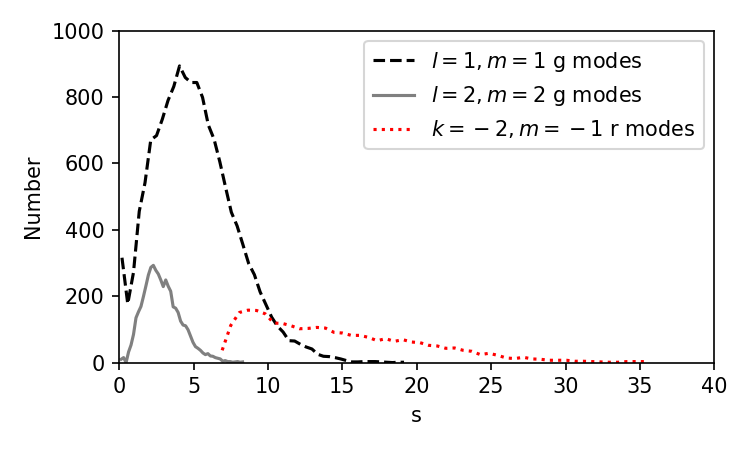}
    \caption{The spin parameter $s$ distributions of $l=1,~m=1$ g modes (black), $l=2,~m=2$ g modes (grey), and $k=-2, m=-1$ r modes (red). Several extremely slowly rotating stars contribute the peak with $s$ near 0 in $l=1,~m=1$ g modes. }
    \label{fig:spin_parameters}
\end{figure}

Using Eq.~\ref{equ:spin_parameter}, we calculate the spin parameters for $l=1,~m=1$ g modes, $l=2,~m=2$ g modes, and $k=-2, m=-1$ r modes. Figure~\ref{fig:spin_parameters} displays their distributions. For $l=1,~m=1$ g modes, the spin parameters show a rapid rise and a slow drop from 0 to 15. Most of the modes have $s$ around 5. Several slow rotators have extremely low spin parameters, which form the peak close to zero. For $l=2,~m=2$ g modes, the spin parameters are lower since the pulsation frequencies are longer than dipole g modes. Most of them are around 2.5. For r modes with $k=-2, m=-1$ , they show different spin parameter distributions. The smallest spin parameter value is $\sim6$ and the highest is $\sim 35$. They show a peak around 9. The r-mode spin parameters are typically larger than that of g modes and have different distributions, implying diverse pulsation properties for them.

\section{Rotation on S--P diagram}\label{sec:rot_on_S_P_diagram}
\subsection{Empirical method to calculate rotation rate}
\begin{figure}
\includegraphics[width=\linewidth]{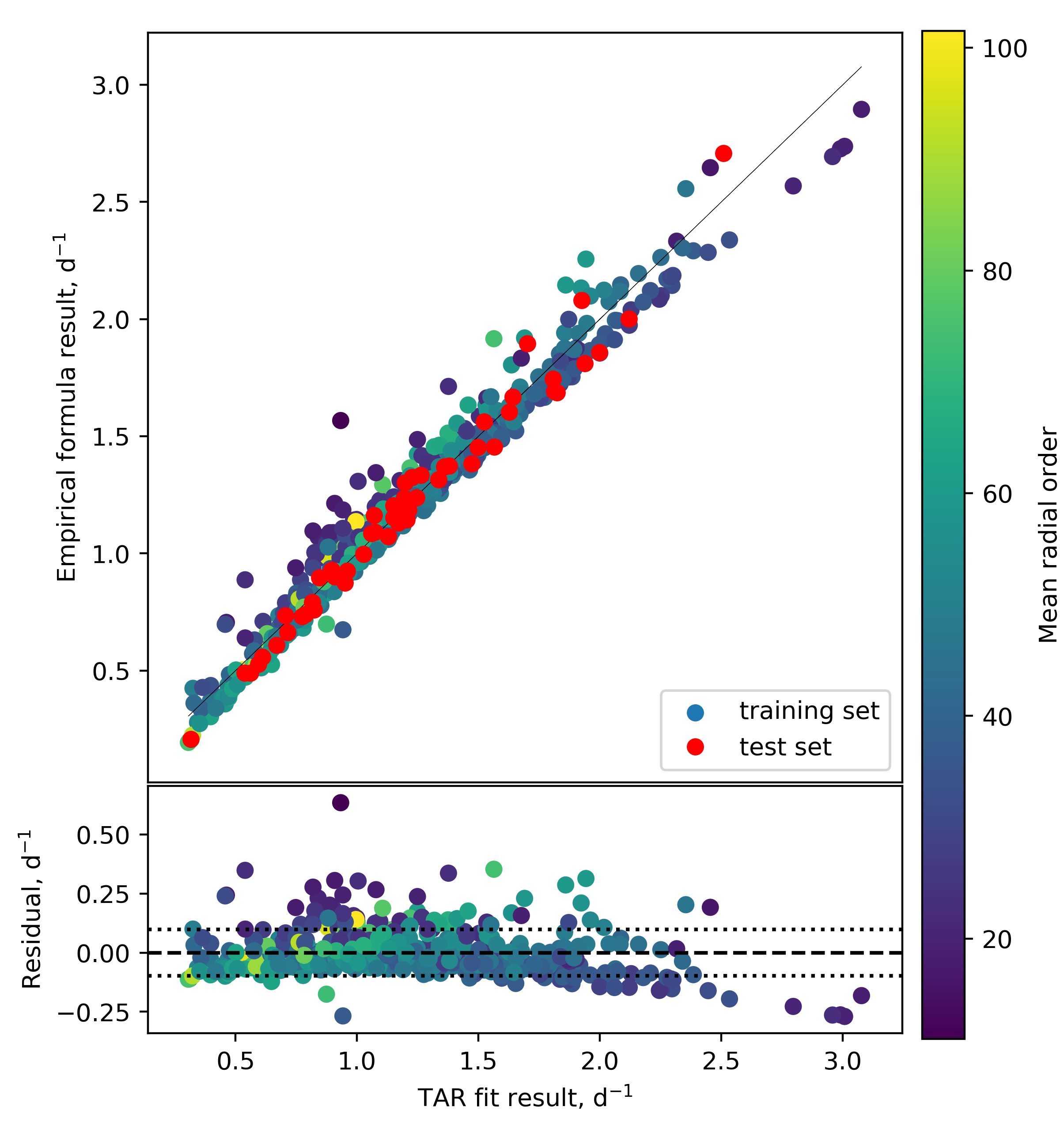}
\caption{The near-core rotation rates calculated by eq.~\ref{equ:empirical} for $k=0, m=1$ g modes. Upper panel: the relation between the input and predicted near-core rotation rates. The red circles are the test set while the others are the training set. Lower panel: the fit residuals. The residuals have the standard deviation of 0.1\,$\mathrm{d^{-1}}$, which is the accuracy of eq.~\ref{equ:empirical}. }\label{fig:empirical_dipole}
\end{figure}



The fit of the TAR reveals the near-core rotation rate, asymptotic spacing, and the estimated radial orders of one pattern, given the quantum numbers $k$ and $m$. This fitting procedure converges faster with a good initial estimate of $f_\mathrm{rot}$. Hence, we report an empirical method to estimate the near-core rotation rate based on the three observables: the mean period $\langle P \rangle$, the mean spacing $\langle \Delta P \rangle$, and the slope $\Sigma$. We use a simple formula to describe the relation between the near-core rotation rate and these three observables. The formula is designed as:
\begin{equation}
f_\mathrm{rot}=\frac{A}{\langle P \rangle} + \frac{B}{\langle \Delta P \rangle} + C\,\Sigma + D, \label{equ:empirical}
\end{equation}
where $A$, $B$, $C$, and $D$ are the coefficients. We selected inversely proportional functions for $\langle P \rangle$ and $\langle \Delta P \rangle$ (both in unit of days) because they are inversely correlated with the rotation rate. For the slope $\Sigma$, the proportional relation is used since rapid rotation causes a steeper period-spacing pattern. On the left hand side of eq.~\ref{equ:empirical}, the unit of \frot is $\mathrm{d^{-1}}$. On the right hand side, the coefficients $A$ and $B$ are dimensionless, the unit of coefficient $C$ and $D$ are $\mathrm{d^{-1}}$. We applied eq.~\ref{equ:empirical} to $l=1,~m=1$ g modes, $l=2,~m=2$ g modes, and $k=-2, m=-1$ r modes, respectively. The slow rotators with $f_\mathrm{rot}<0.4\,\mathrm{d^{-1}}$ were excluded, since the slope is affected by the glitches more than the rotational effect. The best-fitting coefficients are listed in Table~\ref{tab:coefficients}.

\begin{table}
\centering
\scriptsize
\caption{The coefficients of eq.~\ref{equ:empirical} for $k=0, m=1$ g mode, $k=0, m=2$ g mode, and $k=-2, m=-1$ r mode. Note that $k=l-|m|$ for g modes. $A$, $B$, $C$, and $D$ are the coefficients. $\delta f_\mathrm{rot}$ is the fitting accuracy in unit of $\mathrm{d^{-1}}$. }\label{tab:coefficients}
\begin{tabular}{rrrrrr}
\hline
$(k,~m)$  &  $A$    & $B$       &   $C$, $\mathrm{d^{-1}}$  &   $D$, $\mathrm{d^{-1}}$ & $\delta f_\mathrm{rot}$, $\mathrm{d^{-1}}$ \\
\hline
$(0,~1)$ & 0.4189 & 0.001603 &   -11.75      & -0.3554 & 0.1 \\
$(0,~2)$ & 0.3346 & 0.0003965 & -2.477  & -0.2462 & 0.07\\
$(-2,-1)$ & 1.167  & -0.0002585 & -0.2360 & 0.1099 & 0.03\\
\hline
\end{tabular}
\end{table}

Figure~\ref{fig:empirical_dipole} shows the fit result of $k=0, m=1$ g modes. The upper panel reveals the correlation of near-core rotation rates from the TAR fit and the empirical formula eq.~\ref{equ:empirical}. We selected 90\% of the data points (coloured by their mean radial orders) as the training set to obtain the coefficients (listed in the first line of Table~\ref{tab:coefficients}), and use the other 10\% (red circles) to test if the coefficients work well and to avoid over-fitting. Both the training set and the test set show a positive correlation, which means that eq.~\ref{equ:empirical} with the parameters in Table~\ref{tab:coefficients} can estimate the near-core rotation rate. The lower panel shows the differences between the input and output rotation rates. The differences have a standard deviation of $0.1$\,$\mathrm{d^{-1}}$, which is the precision of eq.~\ref{equ:empirical} for $k=0, m=1$ g modes. We find that the mean radial orders show a gradient, in the sense that the points with small residuals have larger mean radial orders than those with large differences. Hence the scatter of eq.~\ref{equ:empirical} is partially caused by the mean radial order. 

The second and third lines in Table~\ref{tab:coefficients} list the coefficients of eq.~\ref{equ:empirical} but for $k=0, m=2$ g modes and $k=-2, m=-1$ r modes. The precision of eq.~\ref{equ:empirical} for $k=0, m=2$ g modes is 0.07\,$\mathrm{d^{-1}}$, which is similar to the $k=0, m=1$ g-mode residuals. However, The precision of $k=-2, m=-1$ r modes is 0.03\,$\mathrm{d^{-1}}$, significantly smaller that those of g modes. The reason is that r modes are not affected by the range of radial orders as much as g modes.

Equation~\ref{equ:empirical} with the coefficients in Table~\ref{tab:coefficients} gives the relation between the rotation rate and the three observables. They can be used as the estimate of the near-core rotation rate before running the TAR fit code. We also tried to search the formula for asymptotic spacing $\Pi_0$ and mean radial order as a function of $\langle P \rangle$, $\langle \Delta P \rangle$ and $\Sigma$, but there are no clear correlations.

\subsection{Rotation on the S--P diagram}

\begin{figure*}
\centering
\includegraphics[width=1\linewidth]{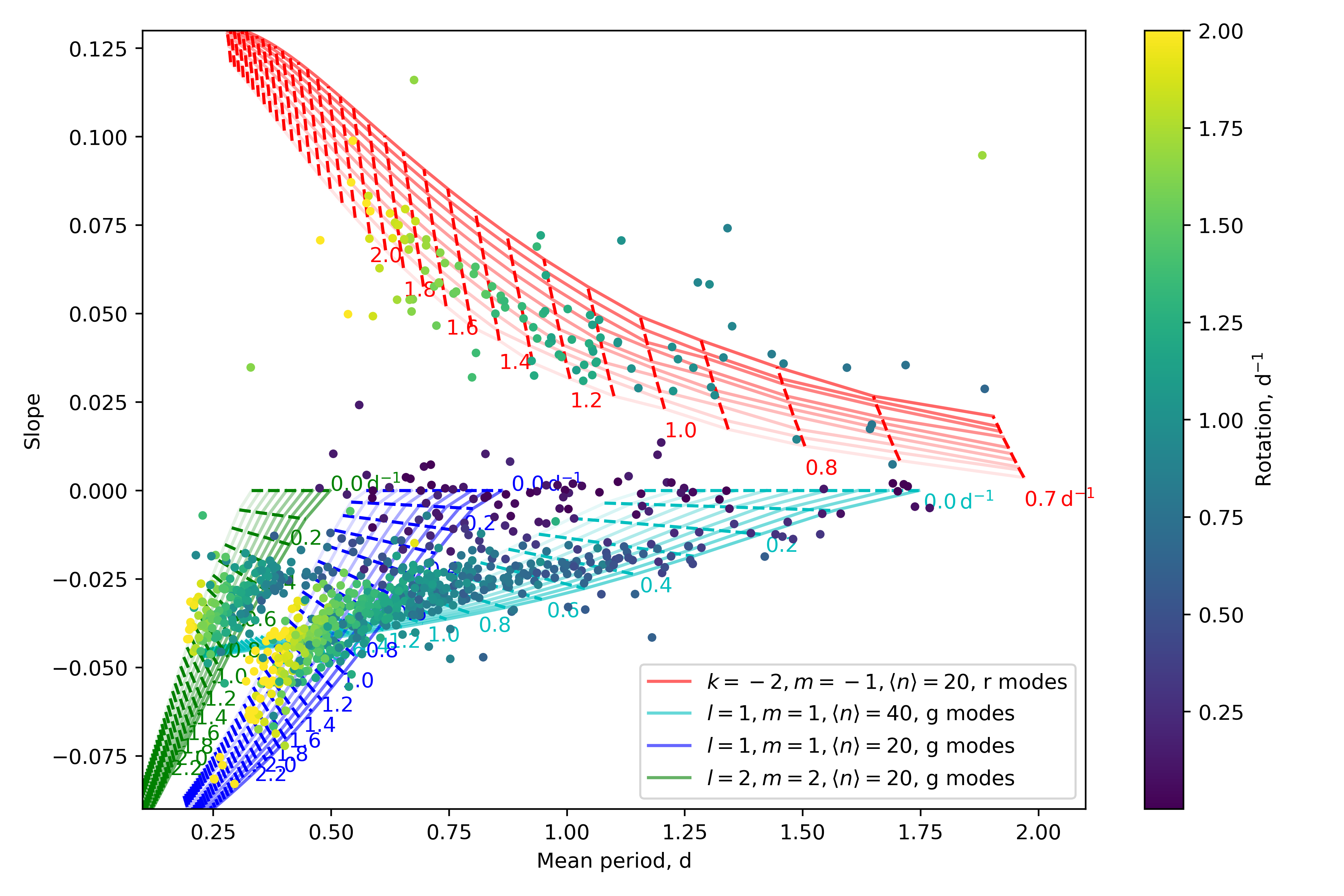}
\caption{Theoretical S--P diagram. One solid line has the same quantum numbers, asymptotic spacing, and radial order centres with rotation increasing from right to left. The dashed lines connect the positions with same rotation rate. Different colours show the curves with different quantum numbers. The transparency stands for the asymptotic spacing $\Pi_0$. The lighter, the smaller the $\Pi_0$. The numbers show the rotation rates in $\mathrm{d^{-1}}$ of the nearest dashed lines. The parameters for those curves are listed in Table~\ref{tab:parameters}. The data points are coloured by their near-core rotation rates to make the comparison with the simulated rotation straightforward.  }\label{fig:iso_rotation}
\end{figure*}

\begin{table}
\centering
\caption{The parameters of the simulated curvatures in Fig.~\ref{fig:iso_rotation}. $k$ and $m$ are the quantum numbers of the modes. All the patterns have 20 modes with different radial order centres $\langle n \rangle$. $\Pi_0$ is the asymptotic spacing and $f_\mathrm{rot}$ is the near-core rotation rate. }\label{tab:parameters}
\begin{tabular}{rrrrrr}
\hline
Colour & $\left(k,m\right)$    & $\langle n \rangle$ & $\Pi_0$,\,s             & $f_\mathrm{rot}$,\,\,$\mathrm{d^{-1}}$      \\
       &                       &                     &  min, max, step        &  min, max, step \\
\hline
Blue   & $(0,1)$    & 20  & 3600, 5600, 200   & 0.0, 4.0, 0.1     \\
Cyan   & $(0,1)$    & 40  & 3600, 5600, 200   & 0.0, 4.0, 0.1      \\
Green  & $(0,2)$    & 20  & 3600, 5600, 200   & 0.0, 4.0, 0.1      \\
Red    & $(-2,-1)$  & 20  & 3600, 5600, 200   & 0.7, 4.0, 0.1     \\
\hline
\end{tabular}
\end{table}

To plot the rotation rate on the S--P diagram, we used the TAR to simulate the period-spacing pattern and calculate the synthetic mean periods and slopes. We show our simulated curves in Fig.~\ref{fig:iso_rotation}, whose parameters are listed in Table~\ref{tab:parameters}. For both g and r modes, the number of modes in each pattern was selected as 20 and the range of $\Pi_0$ is from 3600\,s to 5600\,s with step of 200\,s. For g modes, the rotation is between 0 to 4.0\,$\mathrm{d^{-1}}$ with step of 0.1\,$\mathrm{d^{-1}}$ to cover the observed range. Two regions of radial orders were used for g modes, centred at 20 ($10<n<30$) and 40 ($30<n<50$), as shown in blue and cyan lines in Fig.~\ref{fig:iso_rotation}. For r modes, only radial orders around 20 ($10<n<30$) and $0.7\,\mathrm{d^{-1}}<f_\mathrm{rot}<4.0\,\mathrm{d^{-1}}$ are displayed because these parameters regions can explain the r modes data well. 

Figure~\ref{fig:iso_rotation} shows the simulated results of the S--P diagram. We find that the simulated curves cover the data points well. For $l=1,~m=1$ g modes, the data show two trends: one is the patterns with lower radial orders (the blue curves) which show shorter mean periods and the steeper relations between the slope and the period; another trend shows higher radial orders (cyan curves) whose mean periods are generally longer and the relation between the slope and the period is flatter. Two trends have an overlap over $P\sim0.5\,\mathrm{d}$. 
For $l=2,~m=2$ g modes, the trend between the slope and the mean period is not obvious due to the limited detection of quadrupole g modes. So only $10<n<30$ is used to cover $l=2,~m=2$ g modes.

The S--P diagram is a map for the near-core rotation rate, as marked by the dashed lines and numbers in Fig.~\ref{fig:iso_rotation}. The dashed lines connect the positions with same rotation rates, hence we can estimate the near-core rotation rate by placing the star on the S--P diagram. The rapidly rotating stars generally appear on the left in the S--P diagram, because of both the Coriolis force and the transformation between the co-rotating and inertial reference frame. The estimate of the near-core rotation rate is affected by the mode identification, the asymptotic spacing, and the radial orders. There is an overlapping area around $P\sim0.5\,\mathrm{d}$. In this area, the pattern with higher radial order (cyan curves) shows a higher rotation rate ($\sim1.6\,\mathrm{d^{-1}}$) while the pattern with low radial order (blue curves) has a slower rotation rate ($\sim1\,\mathrm{d^{-1}}$). 
For r modes, the relation is clearer and the S--P map can give a better estimate for the near-core rotation rate. It also explains why the r-mode residuals in Table~\ref{tab:coefficients} are smaller than those for the g-mode, since there is only one trend for r modes on the S--P diagram.

\section{Fast rotators with splittings}\label{sec:fast_splitting}
\begin{figure*}
    \centering
    \includegraphics[width=1\linewidth]{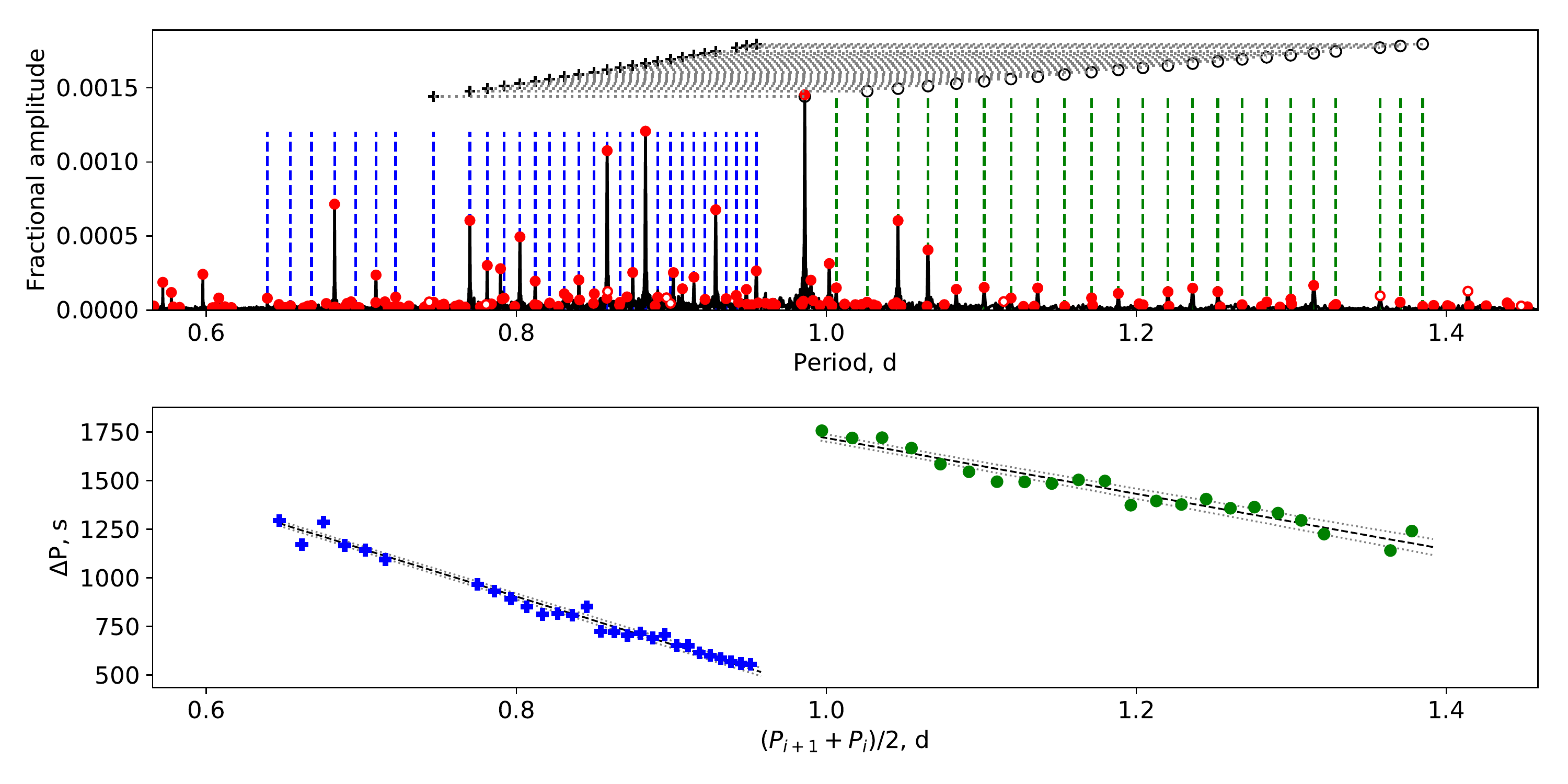}
    \caption{The amplitude spectrum and period-spacing patterns of KIC\,7701947. Top panel: the amplitude spectrum as a function of period. The red dots are the extracted frequencies and the open dots show the likely combination frequencies. The vertical dashed lines show the linear fit by Eq.~\ref{equ:P_i}. Two period-spacing patterns are seen. The left one denoted by the blue dashed lines is the $l=1,~m=1$ g modes. The right one (green vertical lines) is the $l=1,~m=0$ g modes. Above the spectrum `+' is the $m=1$ modes and `O' is the $m=0$ modes. The horizontal dotted lines connect the modes with same radial order $n$. Bottom: the period spacing patterns. The left one is the $l=1,~m=1$ g modes while the right one is $l=1,~m=0$ g modes. The linear fits and uncertainties are shown by the black and grey dashed lines, with dips removed. }
    \label{fig:KIC7701947_power_spectrum_splitting}
\end{figure*}

\begin{figure}
    \centering
    \includegraphics[width=1\linewidth]{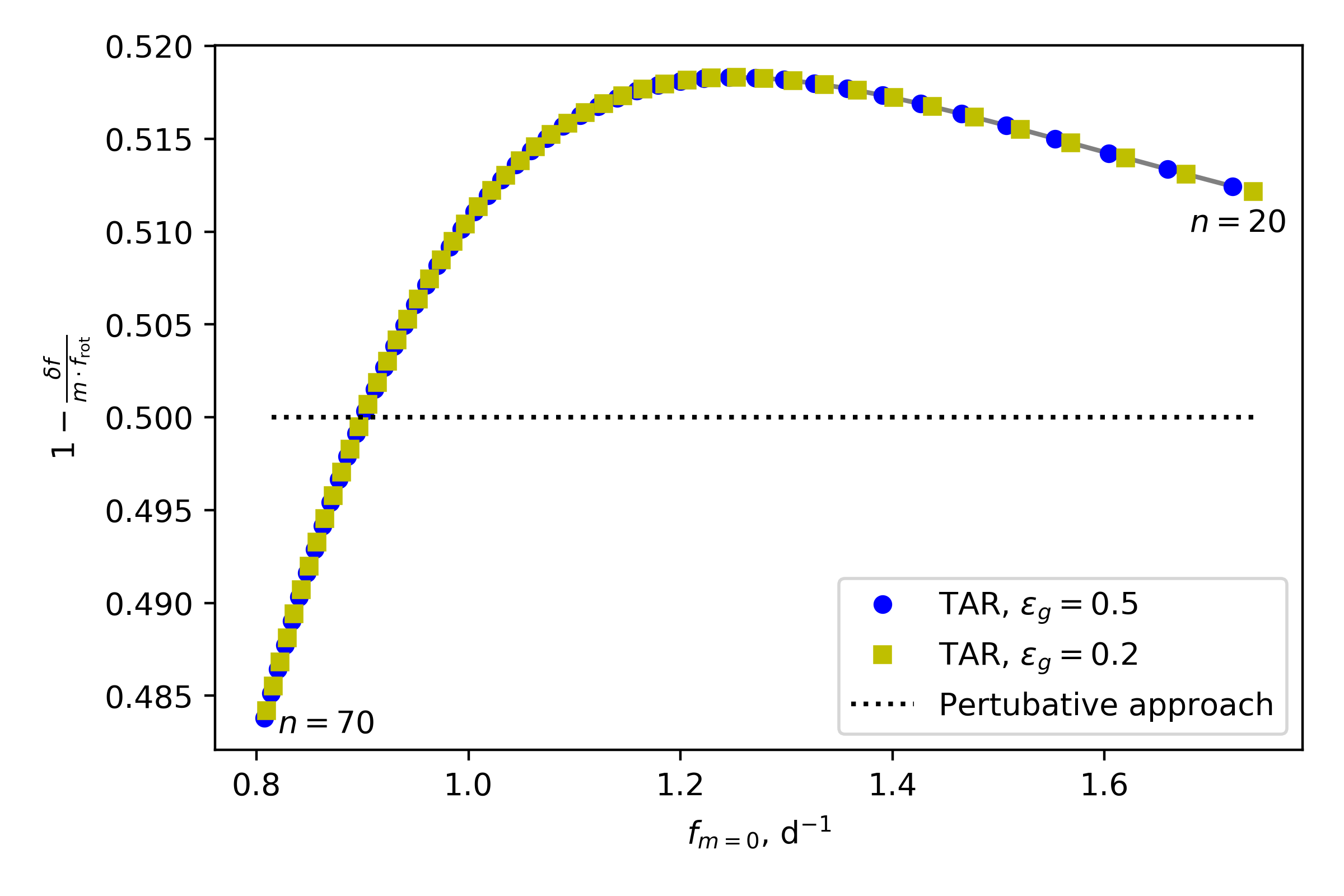}
    \caption{The theoretical Ledoux `constant' as a function of frequency from the TAR, with $\Pi_0=4000\,\mathrm{s}$ and $f_\mathrm{rot}=1\,\mathrm{d^{-1}}$. The considered radial orders have values between 20 and 70. The blue circles and yellow squares show the `constant' with different $\varepsilon_g$. The horizontal dotted line denotes the position with $C_{n,l}=0.5$.}
    \label{fig:theory_splitting}
\end{figure}

\begin{figure}
    \centering
    \includegraphics[width=\linewidth]{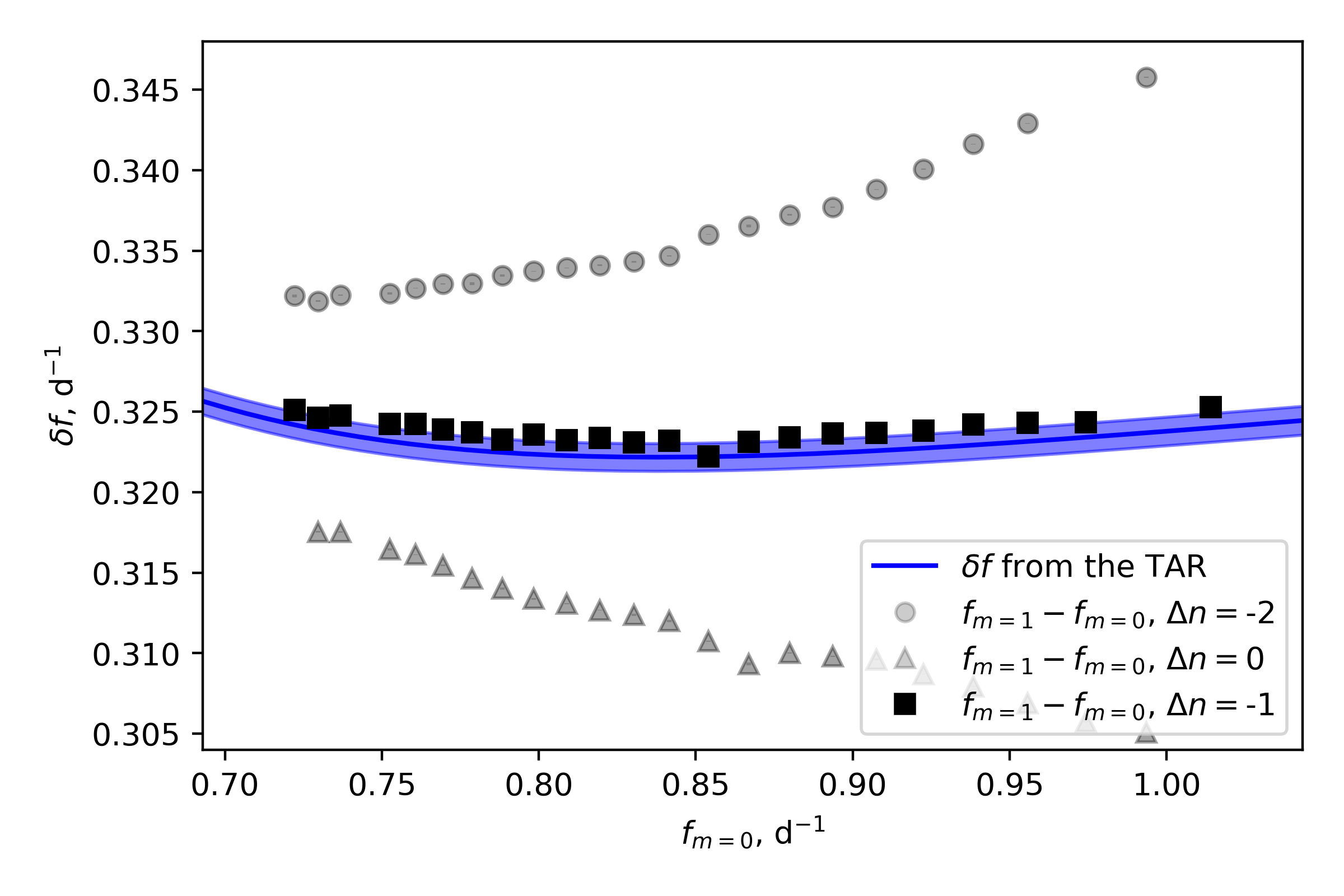}
    \caption{The splitting variation as a function of frequency of KIC\,7701947. The blue line and shaded area are the predicted curve and uncertainty from the TAR fitting result. The black squares are the splittings which follow the theory best, whose radial orders are subtracted by 1 artificially for the best-fitting result. The grey circles and triangles are the splittings whose radial orders are mismatched by a factor of $\pm 1$ around $-1$. }
    \label{fig:KIC7701947_splitting}
\end{figure}

\begin{figure}
\centering
\includegraphics[width=1\linewidth]{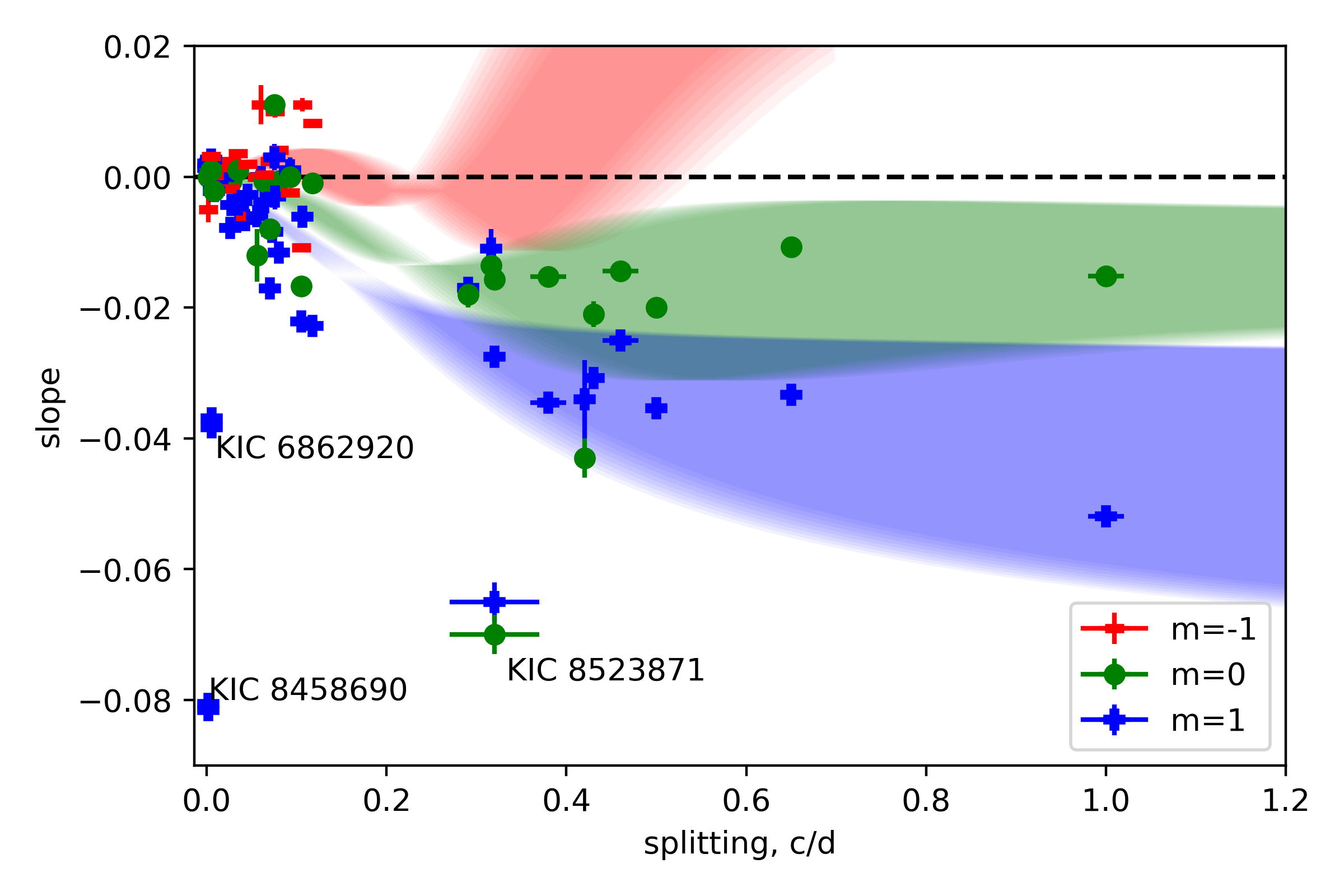}
\caption{Observed relation between slope and splitting from 34 stars. The shaded areas are theoretical curves. We consider the radial orders from 20 to 70 to make the theoretical areas. KIC\,6862920 and KIC\,8458690 are binaries, KIC\,8523871 shows a much larger asymptotic spacing, hence they become outliers. }\label{fig:slope_vs_splitting}
\end{figure}

\begin{table}
    \centering
    \caption{The slopes and splittings of \numberfastsplitting rapidly-rotating stars. $\delta \nu$ are the mean splittings. $\Sigma_{l,m}$ are the slopes. The uncertainty on the last digit is given between brackets. }
    \begin{tabular}{rrrr}
    \hline
    KIC  & $\delta \nu$, $\mathrm{d}^{-1}$ & $\Sigma_{l=1,~m=1}$ & $\Sigma_{l=1,~m=0}$ \\
\hline
3348714 & 0.38(2) & -0.0352(2) & -0.0154(4)\\ 
4285040 & 0.46(2) & -0.014(4) & -0.0163(3)\\ 
4846809 & 0.65(1) & -0.036(6) & -0.0087(7)\\ 
4952246 & 0.316(9) & -0.005(5) & -0.011(1)\\ 
5476473 & 1.00(2) & -0.055(5) & -0.0148(8)\\ 
7701947 & 0.32(1) & -0.0281(1) & -0.0165(5)\\ 
7778114 & 0.50(2) & -0.0355(5) & -0.0202(2)\\ 
8523871 & 0.32(5) & -0.075(5) & -0.067(7)\\ 
9595743 & 0.42(1) & -0.044(4) & -0.027(7)\\ 
12102187 & 0.291(6) & -0.017(7) & -0.019(9)\\ 
12401800 & 0.43(1) & -0.0308(8) & -0.021(1)\\ 
\hline
    \end{tabular}
    \label{tab:my_label}
\end{table}

\cite{Li_2018} reported 22 \gdor stars in which rotational splittings were seen. The rotation rates of those stars are generally slow (with splitting smaller than $0.2\,\mathrm{d^{-1}}$), hence their period-spacing patterns with different azimuthal orders $m$ overlap each other. The traditional \'{e}chelle diagram was used to distinguish the patterns and the shift-copy method helped match the modes with equal radial order $n$ \citep[see details in][]{Li_2018}. 

In this work, we found \numberfastsplitting stars whose splittings are much larger. 
Figure~\ref{fig:KIC7701947_power_spectrum_splitting} displays the splittings of KIC\,7701947 as an example. The top panel shows the power spectrum, in which the red dots are the extracted frequencies and the open dots are the likely combination frequencies. We mark the $l=1,~m=1$ g modes as the blue vertical lines and the $l=1,~m=0$ g modes as the green vertical lines. The plus and circle symbols mark the locations of $m=1$ and $m=0$ modes, respectively. The horizontal dashed lines connect the modes with equal $n$. The bottom panel shows two period-spacing patterns, the left one (blue plus) is $l=1,~m=1$ g modes and the right one (green circle) is $l=1,~m=0$ g modes. The period-spacing patterns look similar to those in Fig.~\ref{fig:KIC7694191} but two features expose the difference: 
\begin{itemize}
    \item the ratio of pulsation periods between $m=1$ and $m=0$ modes does not have a factor of two. 
    \item Slopes of $m=1$ and $m=0$ modes are different so the patterns are not parallel.
\end{itemize}
We also state that these splitting stars are not binaries, since we can use the same parameters ($\Pi_0$, $f_\mathrm{rot}$) to fit both the patterns ($m=1$ or $m=0$) in each star.

Under the condition of slow rotation, the dipole ($l=1$) g-mode splitting is calculated based on the first-order perturbation
\begin{equation}
    \delta f=f_{n, l=1,~m=1}-f_{n, l=1,~m=0}=m\left( 1-C_{n, l}\right)f_\mathrm{rot}, \label{equ:splitting}
\end{equation}
where $f$ is the pulsation frequency, $C_{n, l} \simeq 1/\left[ l\left( l+1\right)\right]=0.5$ is the Ledoux constant, $f_\mathrm{rot}$ is the near-core rotation rate \citep{Ledoux_1951}, the term $m$ is one since we consider dipole g modes. We only consider the splitting between $m=1$ and $m=0$ modes since $m=-1$ retrograde modes are absent in the fast rotating stars \citep[see theory in e.g.][]{Saio_2018}. However, the perturbation is broken with increasing rotation rate. For the newly-discovered splitting stars with much faster rotation rates, the splittings vary between different overtones significantly and it is hard to match the modes with the same $n$. 

For the fast rotators, the TAR in Eq.~\ref{equ:TAR_P} is a good approximation. The frequency in the co-rotating frame is 
\begin{equation}
    f_{nlm, \mathrm{co}}=\frac{\sqrt{\lambda_{l, m, s}}}{\Pi_0 \left(n+\varepsilon_g\right)},
\end{equation}
whose variables are same as Eq.~\ref{equ:TAR_P}. The frequency in the inertial frame is
\begin{equation}
    f_{nlm, \mathrm{in}}=f_{nlm, \mathrm{co}}+m f_\mathrm{rot},
\end{equation}
where $f_\mathrm{rot}$ is the near-core rotation rate. Therefore the splittings is calculated by 
\begin{equation}
    \delta f=m f_\mathrm{rot} +\frac{1}{\Pi_0 \left(n+\varepsilon_g\right)}\left( \sqrt{\lambda_{l, m=1, s}}-\sqrt{\lambda_{l, m=0, s'}}   \right).
\end{equation}
Hence the Ledoux `constant' $C=1-\delta f/(mf_\mathrm{rot})$ is no longer a `constant' \citep[see also][]{Keen_2015, Murphy_2016}. Figure~\ref{fig:theory_splitting} shows the theoretical result of the varying Ledoux `constant'. The parameters are $\Pi_0=4000\,\mathrm{s}$ and $f_\mathrm{rot}=1\,\mathrm{d^{-1}}$, which are chosen from the distributions shown in Fig.~\ref{fig:Pi0_hist} and Fig.~\ref{fig:frot_hist}. We find that with increasing pulsation frequency, the Ledoux `constant' increases from $\sim0.485$, reaches the highest value around $0.52$, and drops slowly. The deviation from the first-order perturbation is $\sim \pm0.02$. We also evaluated different values for $\varepsilon_g$ and find that it does not change the curve, but only shifts the frequencies of zonal modes, as shown by the blue circles and yellow squares in Fig.~\ref{fig:theory_splitting}. 

Figure~\ref{fig:KIC7701947_splitting} shows the comparison between the observed splittings and the theoretical predictions of KIC\,7701947. The blue curve and shaded area show the theoretical splittings and uncertainties, whose parameters are from the best-fitting TAR result. We tried to introduce an artificial shift on the radial order of the $l=1,~m=0$ g modes and plot the results as grey circles and triangles in Fig.~\ref{fig:KIC7701947_splitting}. The black squares are the matching that follows the theory best. We find that changing the mode matching by $\Delta n=\pm1$ indeed changes the shape of splitting. The splittings with correct matching generally follow the theoretical prediction. However, the observed splittings are higher than the theoretical one, showing a discrepancy with the theory. It means the best model which fits the period-spacing patterns can only partly explain the splittings. This reflects the limit of the asymptotic formula of the TAR in Eq.~\ref{equ:TAR_P}, which can be improved by performing a full seismic calculation.

\cite{Ouazzani_2017} showed that the near-core rotation can be deduced roughly by the slope. To make an observed relation between rotation and slope, we combined the \numberfastsplitting rapidly rotating stars with splittings and 22 slow rotating stars from \cite{Li_2018} to extend the observed slope--splitting diagram to a larger splitting area. Figure~\ref{fig:slope_vs_splitting} shows the slope--splitting relation with both the observations and theoretical curves. The observed points cluster into two groups: the left one is composed of the slow rotators with splitting smaller than $0.2$\,$\mathrm{d^{-1}}$ and slope close to 0, while the points with splitting larger than $0.2$\,$\mathrm{d^{-1}}$ are the fast rotators. There is a gap over 0.2\,$\mathrm{d^{-1}}$, which corresponds to the boundary between the slow and fast rotators in Section~\ref{subsec:bimodality_frot}. Due to the effect of rotation, the slopes of the fast rotators deviate from zero and become lower than the slow rotators. The slopes of zonal modes (green) are generally flatter than the slopes of dipole sectoral modes (blue), consistent with the theoretical curves by \cite{Ouazzani_2017} and us. The retrograde modes (red) are only seen in the slow rotators. The reasons of the absence of retrograde modes in fast rotators are: (1) the period spacings are around $10^4\,\mathrm{s}$ hence they are hard to detect; (2) the amplitudes of retrograde modes are concentrated to the equator; (3) an additional latitudinal nodal lines appear when $s>1$ hence retrograde modes become tesseral modes. Therefore, no retrograde modes are expected in fast rotators \citep{Saio_2018}. 

The shaded areas are the theoretical regions of slopes as a function of splittings with different modes. We calculated the simulated period-spacing patterns by the TAR, and measured the slopes of them. The asymptotic spacing $\Pi_0$ is from 3500\,s to 4500\,s with step of 100\,s. The radial order is from 20 to 70. Since the period spacing changes quasi-linearly with period, the slope changes as a function of radial order. We measured the slope at the beginning and at the end of each simulated period-spacing pattern. We found that the slope variation at different radial orders is significant, hence the slope--splitting curves in our work show a much larger dispersion than \cite{Ouazzani_2017}, who neglected this effect. Our simulated curves in Fig.~\ref{fig:slope_vs_splitting} cover the data points in the fast rotation area, but have difficulty in the slow rotation area. The slopes of slowly-rotating stars spread much wider than the shaded area, implying that dips dominate the measurement of slopes in slowly-rotating stars.

We only see \numberfastsplitting rapidly rotating stars with splittings among \starnumber \gdor stars in our sample. The lack may reflect the surface amplitude distribution of tesseral modes. With increasing rotation, the mode geometry are concentrated toward the equator, hence the brightness change by pulsations is cancelled out unless the line of sight is almost aligned with the rotation axis of the star \citep{Townsend_2003_amplitude}. Apart from these \numberfastsplitting splitting stars, we also find two stars, KIC\,5092684 and KIC\,5544996, which show $l=2,~m=1$ modes. Precise observations of their projected equatorial velocities $v \sin i$ will reveal their inclinations and allow us to evaluate the theoretical expectations for the amplitude distributions of the pulsation modes.

\section{Surface modulations}\label{sec:surface_rotation}

\begin{table}
    \centering
    \caption{The near-core and surface rotation rates, and their ratios of \numbersurfacemodulation stars which show surface modulation signals. `EB' means the star is an eclipsing binary and `SURF' means the signal is caused by surface modulations. $f_\mathrm{core}$ is the near-core rotation rate and $f_\mathrm{surf}$ is the surface rotation rate. }
    \label{tab:surface_rotation}
    \begin{tabular}{rrrrr}
    \hline
    KIC      & Type & $f_\mathrm{core}$,$\mathrm{d^{-1}}$  &  $f_\mathrm{surf}$,$\mathrm{d^{-1}}$  &  $f_\mathrm{core}/f_\mathrm{surf}$ \\
    \hline
KIC\,2846358 & SURF & 0.755(4) & 0.754(8) & 1.00(1) \\ 
KIC\,3341457 & EB & 1.859(1) & 1.893(7) & 0.982(4) \\ 
KIC\,3440840 & SURF & 0.93(6) & 0.938(7) & 0.99(6) \\ 
KIC\,3967085 & SURF & 0.76(2) & 0.77(1) & 1.00(3) \\ 
KIC\,4171102 & SURF & 0.7579(6) & 0.76(2) & 1.00(2) \\ 
KIC\,4567531 & SURF & 1.018(5) & 0.94(2) & 1.08(2) \\ 
KIC\,4932417 & SURF & 1.285(8) & 1.23(2) & 1.05(2) \\ 
KIC\,4951030 & SURF & 2.53(3) & 2.49(4) & 1.02(2) \\ 
KIC\,5021329 & SURF & 2.02(2) & 1.98(4) & 1.02(2) \\ 
KIC\,5025464 & SURF & 1.57(3) & 1.561(3) & 1.00(2) \\ 
KIC\,5115637 & SURF & 0.714(6) & 0.72(2) & 0.99(2) \\ 
KIC\,5210153 & SURF & 1.051(4) & 1.040(5) & 1.011(6) \\ 
KIC\,5370431 & SURF & 0.618(5) & 0.6122(6) & 1.009(8) \\ 
KIC\,5374279 & SURF & 1.00(2) & 0.873(4) & 1.14(3) \\ 
KIC\,5608334 & SURF & 2.25(1) & 2.241(3) & 1.005(6) \\ 
KIC\,5652678 & SURF & 1.140(5) & 1.163(6) & 0.980(6) \\ 
KIC\,5876187 & SURF & 0.596(5) & 0.584(4) & 1.02(1) \\ 
KIC\,5954264 & SURF & 1.330(9) & 1.329(6) & 1.001(8) \\ 
KIC\,5978913 & SURF & 0.955(5) & 0.950(6) & 1.006(8) \\ 
KIC\,6041803 & EB & 1.53(2) & 1.524(7) & 1.00(1) \\ 
KIC\,6284209 & SURF & 1.48(2) & 1.42(4) & 1.04(3) \\ 
KIC\,6366512 & SURF & 1.17(1) & 1.18(1) & 0.99(1) \\ 
KIC\,6445969 & SURF & 1.263(7) & 1.244(5) & 1.015(7) \\ 
KIC\,6469690 & SURF & 0.566(5) & 0.554(6) & 1.02(1) \\ 
KIC\,6935014 & SURF & 0.789(3) & 0.788(7) & 1.00(1) \\ 
KIC\,7059699 & SURF & 2.30(2) & 2.27(2) & 1.01(1) \\ 
KIC\,7287165 & SURF & 0.95(1) & 0.986(2) & 0.96(1) \\ 
KIC\,7344999 & SURF & 1.40(1) & 1.40(3) & 1.00(2) \\ 
KIC\,7434470 & EB & 1.77(1) & 1.698(1) & 1.044(6) \\ 
KIC\,7596250 & EB & 1.1876(7) & 1.185(4) & 1.003(4) \\ 
KIC\,7620654 & SURF & 1.88(2) & 1.83(4) & 1.03(2) \\ 
KIC\,7621649 & SURF & 0.7745(4) & 0.7802(6) & 0.9928(9) \\ 
KIC\,7840642 & SURF & 1.158(6) & 1.158(6) & 1.000(8) \\ 
KIC\,7968803 & SURF & 1.94(1) & 1.949(7) & 0.997(8) \\ 
KIC\,8180062 & SURF & 0.907(5) & 0.904(9) & 1.00(1) \\ 
KIC\,8197019 & SURF & 1.89(5) & 1.86(4) & 1.01(3) \\ 
KIC\,8264667 & SURF & 0.684(5) & 0.682(2) & 1.002(8) \\ 
KIC\,8264708 & SURF & 1.636(1) & 1.636(1) & 1.000(1) \\ 
KIC\,8293692 & SURF & 1.00(2) & 1.044(9) & 0.95(2) \\ 
KIC\,9573582 & SURF & 0.9495(5) & 0.946(5) & 1.004(6) \\ 
KIC\,9652302 & SURF & 0.9147(6) & 0.910(2) & 1.005(3) \\ 
KIC\,9716350 & SURF & 0.863(3) & 0.864(1) & 0.999(3) \\ 
KIC\,9716563 & SURF & 0.9081(9) & 0.90(2) & 1.01(2) \\ 
KIC\,9847243 & SURF & 0.933(5) & 0.913(9) & 1.02(1) \\ 
KIC\,10347481 & SURF & 1.303(5) & 1.27(2) & 1.03(2) \\ 
KIC\,10423501 & SURF & 0.8420(6) & 0.841(4) & 1.001(4) \\ 
KIC\,10470294 & SURF & 2.00(2) & 1.96(4) & 1.02(2) \\ 
KIC\,10483230 & SURF & 0.9126(4) & 0.918(2) & 0.995(2) \\ 
KIC\,10669515 & SURF & 1.596(7) & 1.567(4) & 1.019(5) \\ 
KIC\,10803371 & EB & 0.98(2) & 1.011(3) & 0.97(2) \\ 
KIC\,11183399 & SURF & 1.89(2) & 1.868(3) & 1.01(1) \\ 
KIC\,11201483 & SURF & 2.28(2) & 2.23(3) & 1.02(2) \\ 
KIC\,11294808 & SURF & 0.802(5) & 0.7887(7) & 1.016(7) \\ 
KIC\,11395936 & EB & 0.944(7) & 0.953(6) & 0.99(1) \\ 
KIC\,11462151 & SURF & 2.00(2) & 1.93(4) & 1.03(2) \\ 
KIC\,11520274 & SURF & 1.03(2) & 1.025(3) & 1.00(2) \\ 
KIC\,11922283 & SURF & 1.088(8) & 1.07(1) & 1.02(1) \\ 
KIC\,12202137 & SURF & 1.96(1) & 1.96(1) & 0.998(8) \\ 
\hline
    \end{tabular}
\end{table}

\begin{figure}
    \centering
    \includegraphics[width=1\linewidth]{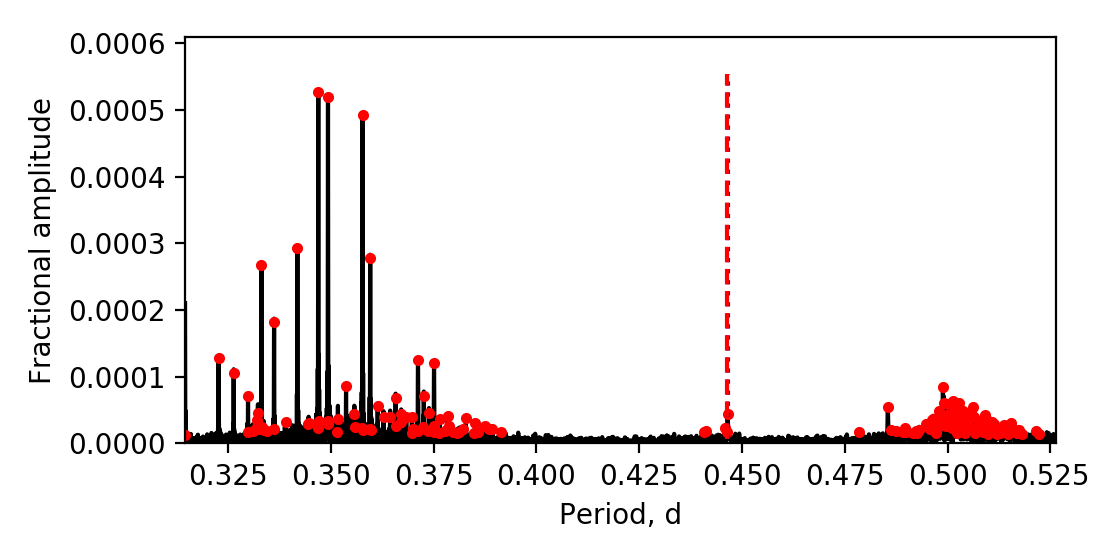}
    \caption{The near-core and surface rotations of KIC\,5608334. The red dots show the frequencies. The red dashed line shows the surface rotation period, which is consistent with the near-core rotation rate derived by the g-mode pattern. The hump at 0.5\,d is the unsolved r modes. }
    \label{fig:KIC5608334_surface_rotation}
\end{figure}

\begin{figure}
    \centering
    \includegraphics[width=1\linewidth]{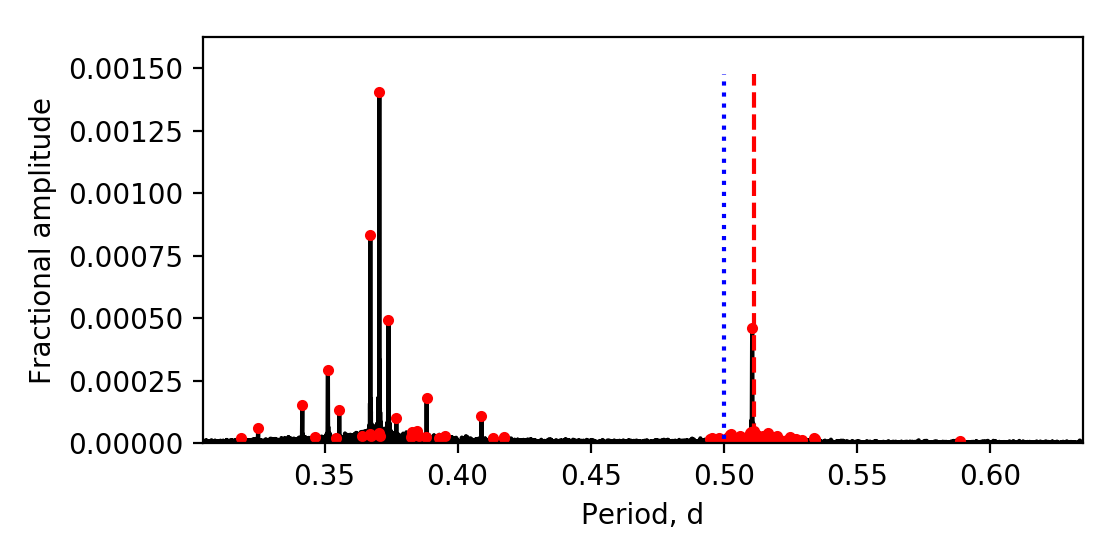}
    \caption{Same as Fig.~\ref{fig:KIC5608334_surface_rotation} but for KIC\,10470294. The red vertical dashed line shows the surface rotation period while the blue dotted line gives the near-core rotation rate. We identify that in this star the hump at 0.52\,d is the surface rotation signal, not r modes. }
    \label{fig:KIC10470294_surface_rotation}
\end{figure}

\begin{figure}
    \centering
    \includegraphics[width=\linewidth]{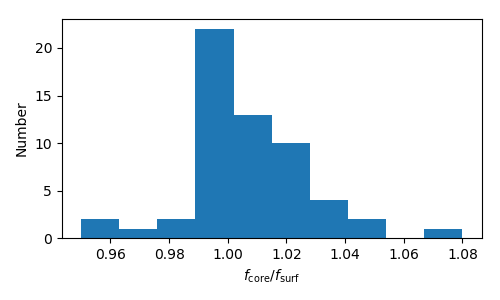}
    \caption{The distributions of the core-to-surface rotation ratio.}
    \label{fig:rotation_ratio_distribution}
\end{figure}


The g and r modes allow us to measure the rotation rate of the near-core region. To see how the rotation rate changes radially, which is crucial to understand the angular momentum transfer, we looked for the surface modulation signals. The surface modulation might be caused by spots \citep[e.g.][]{McQuillan_2014}, or stellar activities, which is straightforward to detect since the signal is located within the typical g mode period region. Although we excluded the eclipsing binaries (EB) when selecting the stars, there are still several stars which were classified as EB by \cite{Kirk_2016}, since their eclipses are too shallow to be seen in the time domain. The orbital periods are equal to the surface rotation periods for short period EBs, considering the components are tidally locked. Hence we can use the orbital period as the surface rotation period. We follow the criteria used by \cite{VanReeth_2018} and \cite{Li_2019} to select the rotational modulation signal. The criteria are:
\begin{itemize}
    \item the surface modulation is a closely-spaced group of peaks in the amplitude spectrum, assuming the lifetime of spot is shorter than the 4-yr observation.
    \item the signal-to-noise ratio (S/N) is larger than 4.
    \item the harmonic of the highest peak is seen.
    \item the signal is located between g and r modes. For those without r modes, we search the signal between one and two times the g-modes mean period, because the mean period of r modes is approximately twice the mean period of g modes.
    \item Both the rotational modulation signal at the rotation frequency and its harmonic have to be well-separated from the period-spacing patterns, to avoid mistaking pulsation modes for rotational modulation.
\end{itemize}

We find \numbersurfacemodulation stars which show surface modulations, representing about 9.5\% of the \starnumber stars. The fraction is consistent with spectroscopic observations of Zeeman splitting \citep{Donati_2009, Wade_2016} and previous photometric observations of smaller samples \citep{VanReeth_2018, Li_2019}. Table~\ref{tab:surface_rotation} lists the KIC numbers of these stars, the near-core ($f_\mathrm{core}$) and surface ($f_\mathrm{surf}$) rotation rates, and their ratios. Figure~\ref{fig:KIC5608334_surface_rotation} gives the amplitude spectrum of KIC\,5608334 which shows a surface rotation signal. We find a peak group at $\sim 0.44$\,d (red dashed line), which is consistent with the near-core rotation rate derived by the g-mode period-spacing pattern. A hump in Fig.~\ref{fig:KIC5608334_surface_rotation} appears around 0.5\,d, which we identify as unresolved r modes \citep[e.g.][]{Saio_2018_Rossby_mode}. 

However, not all such humps are r modes. Figure~\ref{fig:KIC10470294_surface_rotation} displays the surface modulation of KIC\,5608334, in which the hump at $\sim 0.52$\,d lies over the near-core rotation period. This hump is not unresolved r modes since the latter have periods longer than the rotation period. We conclude that this hump is caused by the spots on the surface, as pointed out by e.g. \cite{Balona_2013}. 

We calculate the ratio between the near-core and surface rotation and show them in Fig.~\ref{fig:rotation_ratio_distribution}. We find that almost all the stars rotate nearly rigidly since the ratio is between 0.95 and 1.08. The distribution shows a sharp drop at $\sim0.99$, implying that the near-core region rotates slightly faster than the surface, which is consistent with the theoretical rotation profile reported by \cite{Rieutord_2006}. A possible selection bias needs to be mentioned. A surface rotation period longer than the near-core rotation period might not be detected since this signal may be identified as a peak group of unresolved r modes.



\section{Conclusions}\label{sec:conclusions}

We report \patternnumber period-spacing patterns detected from \starnumber \gdor stars, including 22 slow rotators with rotational splittings, \numberfastsplitting rapid rotators with rotational splittings, \numberossbyrmode stars with r modes, and \numbersurfacemodulation stars that present surface modulation signals. The majority (62.0\%) of the detected modes are $l=1,~m=1$ g modes. We also see many $l=2,~m=2$ g modes and $k=-2, m=-1$ Rossby modes, which comprise 18.9\% and 11.5\% of the sample, respectively. Among the \starnumber \gdor stars, there are 339 stars which only show dipole g modes, 145 stars showing both dipole and quadrupole g modes, 83 stars showing dipole g modes and $k=-2, m=-1$ r modes, 27 stars showing dipole, quadrupole g modes, and r modes. We also find 16 stars whose dipole g modes cannot be resolved, and one star which does not show any dipole g mode power.

For each pattern, a series of pulsation periods were identified. The mean periods, the mean period spacings, and the slope were calculated for each pattern. We find \gdor stars have a relation on the slope--mean period diagram (S--P diagram, Fig.~\ref{fig:period_slope}), where the data points cluster into different areas based on their quantum numbers. The S--P diagram gives the typical pulsation period and slopes of \gdor stars. For $l=1,~m=1$ g modes, the periods are between 0.3\,d and 1.2\,d with a slope around $-0.03$, while for $l=2,~m=2$ g modes, the period is half that of the $l=1,~m=1$ g modes but the slope is similar. Both $l=2,~m=2$ and $l=1,~m=1$ g modes show a positive correlation between the slope and the mean period, which is an effect of rotation confirmed by the TAR simulations.

We obtain the near-core rotation rates, the asymptotic spacings, and the radial orders using the TAR. We find that the distribution of the near-core rotation rate shows a slow-rotator excess, similar to the previous observations of the projected velocity $v \sin i$. There are more slow rotators than the angular momentum transfer models by \cite{Ouazzani_2018} predicted, implying some additional mechanisms of angular momentum transfer are present inside these stars, or the effect of overshooting is significant. We obtained \numberfastsplitting fast rotators that show splittings, whose modes are $l=1,~m=1$ and $l=1,~m=0$ g modes. Due to the rapid rotation, the splitting varies as a function of radial order. We find that the best-fitting TAR result can explain the period-spacing patterns but it can only partly explain the splittings. Surface modulations are found in \numbersurfacemodulation stars, with rotation rates close to the near-core rotation rates. Most \gdor stars rotate rigidly, with the near-core region rotating slightly faster, but not by more than 5\%.

Our observational sample is large enough to identify some outstanding problems in theoretical models of \gdor stars:
\begin{itemize}
\item \textcolor{black}{Most stars in the \gdor instability strip do not show period spacing patterns, or their patterns are incomplete. Mode selection mechanisms for \gdor pulsations are needed. }
\item \textcolor{black}{Concerning \gdor pulsation excitation, we confirm that a number of \gdor stars have hotter temperatures and also excite more radial orders than theoretically predicted.}
\item \textcolor{black}{We observe that the near-core regions of \gdor stars rotate more slowely than expected, in disagreement with the theory. Two directions might be considered: fast rotation hinders g-mode pulsations, or our model for angular momentum transport is missing some key mechanisms.}
\end{itemize}

\section*{Acknowledgements}
The research was supported by an Australian Research Council DP grant DP150104667. Funding for the Stellar Astrophysics Centre is provided by the Danish National Research Foundation (Grant agreement no.: DNRF106). TVR has received funding from the European Research Council (ERC) under the European Union's Horizon 2020 research and innovation programme (grant agreement N$^\circ$670519: MAMSIE) and from the KU\,Leuven Research Council (grant C16/18/005: PARADISE). 
This work has made use of data from the European Space Agency (ESA) mission
{\it Gaia} (\url{https://www.cosmos.esa.int/gaia}), processed by the {\it Gaia}
Data Processing and Analysis Consortium (DPAC,
\url{https://www.cosmos.esa.int/web/gaia/dpac/consortium}). Funding for the DPAC
has been provided by national institutions, in particular the institutions
participating in the {\it Gaia} Multilateral Agreement.




\bibliographystyle{mnras}
\bibliography{ligangref} 




\appendix
\newpage
We present four appendices: period-spacing patterns and TAR fittings, parameters of the \gdor stars, \numberfastsplitting rapidly rotating stars with rotational splittings, and \numbersurfacemodulation stars with surface rotation signals. The appendices are online only, here we only give the descriptions.

\section{Period-spacing patterns and TAR fits}\label{appendix:period_spacing_patterns}
We present the periodograms and period spacing patterns of \starnumber $\gamma$\,Dor stars. For each star, we show the periodogram with identified modes, the period spacing pattern(s) and the linear fitting(s), and sideways \'{e}chelle diagram(s). We also show the TAR fitting and the posterior distributions of the near-core rotation rate and the asymptotic spacing. 

\section{Parameters of the \gdor stars}\label{appendix:long_table}
We list the observed and TAR fitting parameters of \starnumber \gdor stars in this paper. The parameters are: the \textit{Kepler} magnitudes, the effective temperatures, the luminosities, the mode identifications ($k\equiv l-|m|$, $l$ is the angular degree and $m$ is the azimuthal order), mean pulsation periods $\langle P \rangle$, mean period spacings $\langle \Delta P \rangle$, slopes $\langle \Sigma \rangle$, asymptotic spacings $\Pi_0$, near-core rotation rates $f_\mathrm{rot}$, the ranges of radial orders $n$, and ranges of spin parameters $s$. \textcolor{black}{We also mark the stars which have short-cadence data or have p modes oscillations.} The full version of this table can be found online.

\section{Rapidly-rotating stars with rotational splittings}\label{appendix:rapidly_splitting}
We find \numberfastsplitting stars that rotate rapidly but still show rotational splittings. These stars are interesting because rotational splittings are rare among rapid rotators. The inclinations of these stars should be very small so the tesseral modes are seen, based on the amplitude distribution theory.

\section{Surface modulations of \numbersurfacemodulation stars}
\numbersurfacemodulation stars show surface modulation signals. Most of them are caused by the surface activities while a few are classified as eclipsing binaries. Their orbital periods are short hence we assume their surfaces have been tidally locked.

\bsp	
\label{lastpage}
\end{document}